\documentclass[11pt]{article}
\usepackage{newinutile}
\usepackage{latexsym,amsmath,amsfonts,amsthm,amssymb}
\usepackage{epsfig,graphics,color,calc,graphicx,pict2e}
\usepackage{multirow}
\usepackage{cite}
\usepackage{mathrsfs}
\usepackage{tikz}
\usepackage{tikz}
\usetikzlibrary{shapes,arrows}
\usetikzlibrary{intersections,matrix, positioning}

\usepackage{subfigure} 

\newcommand{\newatop}[2]{\genfrac{}{}{0pt}{}{#1}{#2}}

\newlength{\pecettawidth}
\begin{document}
\title{Metastability of synchronous and asynchronous dynamics}

\author{Emilio N.M.\ Cirillo}
\email{emilio.cirillo@uniroma1.it}
\affiliation{Dipartimento di Scienze di Base e Applicate per l'Ingegneria, 
             Sapienza Universit\`a di Roma, 
             via A.\ Scarpa 16, I--00161, Roma, Italy.}

\author{Vanessa Jacquier}
\email{vanessa.jacquier@unifi.it}
\affiliation{Dipartimento di Matematica e Informatica ``Ulisse Dini'', 
viale Morgagni 67/a, 50134, Firenze, Italy.}

\author{Cristian Spitoni}
\email{C.Spitoni@uu.nl}
\affiliation{Institute of Mathematics,
University of Utrecht, Budapestlaan 6, 3584 CD Utrecht, The~Netherlands.}


\begin{abstract}
Metastability is an ubiquitous phenomenon in nature, which interests 
several fields of natural sciences.
Since metastability is a genuine non--equilibrium phenomenon,
its description in the framework of thermodynamics and statistical mechanics
has progressed slowly for a long time.
Since the publication of the first seminal paper in which 
the metastable behavior of the mean field Curie--Weiss model was 
approached by means of stochastic techniques, this topic has been 
largely studied by the scientific community. 
Several papers and books 
have been published in which many different spin models were 
studied and different approaches were developed. 
In this review we focus on the comparison between the 
metastable behavior of synchronous and asynchronous dynamics, 
namely, stochastic processes in discrete time in which at each time 
either all the spins or one single spin are updated. 
In particular we discuss how the two different stochastic 
implementation of the very same Hamiltonian give rise to 
different metastable behaviors.
\end{abstract}





\vfill\eject

\maketitle

\section{Introduction}
\label{s:intro} 
\par\noindent
Metastable states are commonly 
observed in several diverse fields such as 
physics, 
chemistry, 
biology, 
computer science, 
climatology, 
and economics.
In order to portray 
their main features 
we use, here, the classical example of 
super--saturated vapors: 
under special 
experimental conditions,
a vapor can be compressed at pressures lower than the value 
at which liquefaction should start. The vapor, thus, enters a
state different from the thermodynamic equilibrium phase, 
but, for small variations of the thermodynamics parameters, it 
behaves as if it were in real equilibrium following the laws of thermodynamics
and undergoing small reversible changes. 
The vapor can remain in such a state for a very long time, but it 
can exit such a state, and reach the real thermodynamic phase (the liquid
phase), via internal random fluctuation or external perturbations
\cite{penrose1987}.
The eventual transition from the vapor to the liquid phase is 
irreversible. 

Other well known contexts in which metastability is observed 
are crystallization of proteins
and ferromagnetic materials (branches of the hysteresis loop 
where the magnetization is opposite to the external magnetic field).

The study of metastability has attracted much attention in the last
decades not only for its intrinsic interest, but also because 
metastable states are an example of non--equilibrium states of 
thermodynamical systems. As it is well known, Statistical Mechanics 
has developed a complete mathematical formalism, even on 
rigorous basis \cite{Ruelle}, to describe equilibrium states, whereas
a complete consistent theory of systems out of equilibrium is 
still lacking. Thus, deriving a rigorous mathematical theory of 
metastable states can shed same light on rigorous approaches to 
non--equilibrium thermodynamics.

\begin{figure}
\begin{center}
\includegraphics[width=0.45\textwidth]{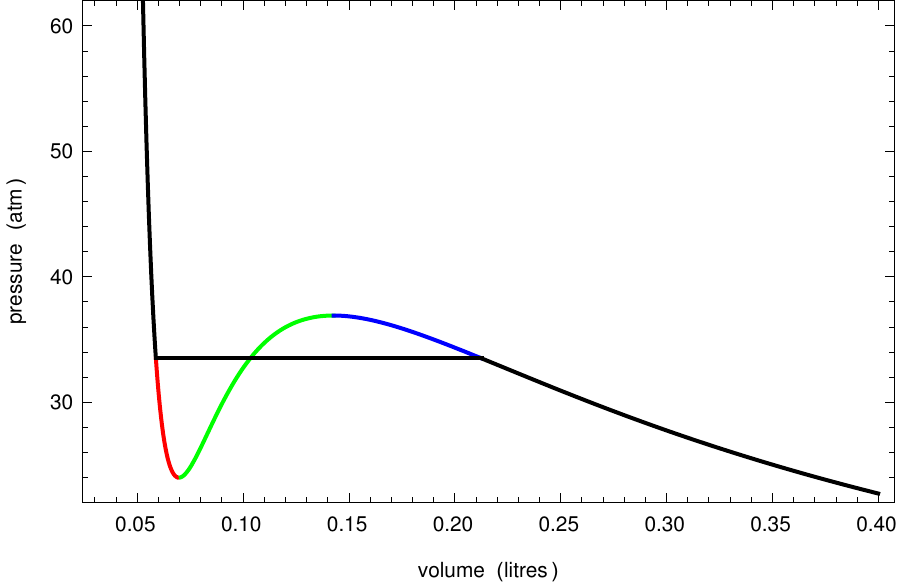}
\hskip 1 cm
\includegraphics[width=0.45\textwidth]{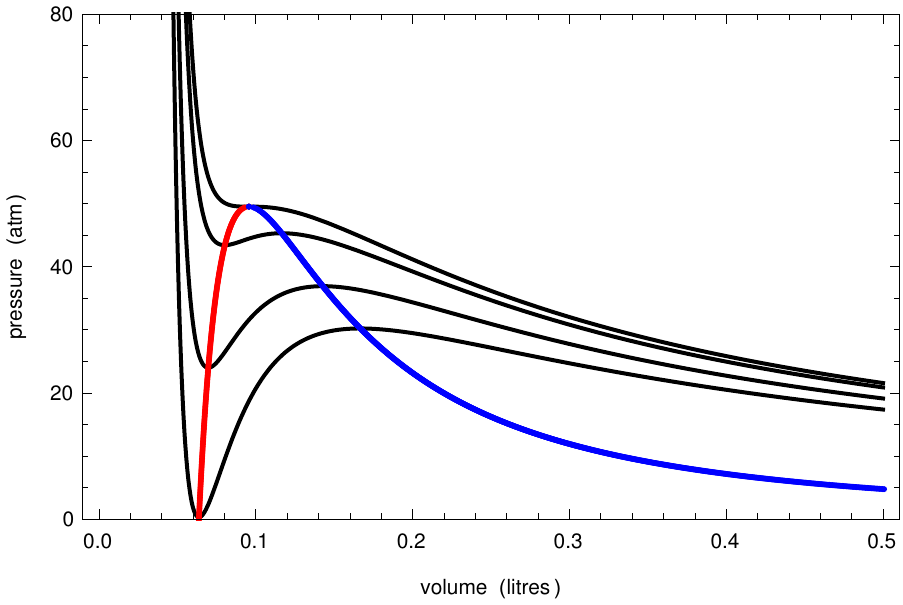}
\end{center}
\caption{Left: the Maxwell equal area rule is used to replace 
part of the van der Waals isotherm with an horizontal segment yielding
the isotherm describing the liquid--vapor transition (black curve).
The van der Waals isotherm 
$(P+a/V^2)(V-b)=RT$, has been plotted for
$a=1.36~\ell^2$ atm mol$^{-2}$ and
$b=0.0319~\ell$ mol$^{-1}$, 
at $T=140^\textup{o}$K, with 
$R=0.082058~\ell$ atm mol$^{-1}$. 
The equal area rule yields
$33.48$~atm as the value of the pressure at which 
vapor and liquid coexists.
Right:
black lines, from the bottom to the top,
are van der Waals isotherms for the same $a$ and $b$
at temperatures $T=130,140,150,153.94^\textup{o}$K.
The red and the blue curves are, respectively, the liquid and the 
vapor branches of the spinodal curve, which meet at the critical 
point 
$V_\textup{crit}=3b=0.0957~\ell$
and
$P_\textup{crit}=a/(27b^2)=49.50$~atm 
on the critical isotherm at 
$T_\textup{crit}=8a/(27Rb)=153.94^\textup{o}$~K.
Note that the values of $a$ and $b$ that we used are the ones valid for
the Oxygen, the experimental value of the critical temperature 
is $156^\textup{o}$K and the vapor pressure at temperature 
$140^\textup{o}$K is $27.50$~atm \cite{Weber}.
}
\label{f:fig01}
\end{figure}

The first attempts to provide a theoretical explanation of metastability 
are based on the real gas van der Waals equation \cite{vanderWaals},
which, coupled with the Maxwell construction
\cite{Maxwell,Maxwell-libro}, can be interpreted as an 
equation describing the liquid--vapor transition (see the 
left panel of figure~\ref{f:fig01}). 
Below the critical temperature 
the van der Waals isotherms
have a not monotonic behavior, but, using the \emph{equal area}
Maxwell rule the kink can be replaced by a segment joining the 
high pressure part of the curve representing the liquid and the 
low pressure part of the curve representing the vapor. 

At the points in the part of the original isotherm 
curve between the minimum and the maximum, called the 
\emph{instability} branch of the isotherm, 
the compressibility would be negative, 
since $\partial v/\partial p>0$ (see the green part of the isotherm in the
left panel of figure~\ref{f:fig01}).
The locus of the minimum and maximum points of the sub--critical 
isotherms in the plane $V$--$P$ is called the \emph{spinodal curve}:
the minimum and the maximum points form, respectively,
the liquid and the vapor branches of the curve
(red and blue curves in the
right panel of figure~\ref{f:fig01}). 
All the points below the
spinodal curve cannot represent equilibrium states of the 
real gas since they follow on the instability branch of one of 
the van der Waals isotherm and, thus, are mechanically unstable since 
they would have 
negative compressibility. Indeed, if a real gas is prepared in 
one of those states the liquid and the vapor phase quickly separate 
through a mechanism called \emph{spinodal} decomposition. 
On the other hand, if the real gas is prepared in states above the 
spinodal curve not belonging to the pure phase branches of the Maxwell 
isotherms, under particular experimental conditions, it is possible
to observe metastable 
states, such as super--saturated vapor and super--heated liquid.
It is thus rather natural to interpret those points 
(respectively,
the red and the blue arcs in the left panel of figure~\ref{f:fig01})
of the van der Walls isotherms as metastable states. 

Although several important studies tried to develop rigorous theories
of metastability in the framework of thermodynamics and 
Statistical Mechanics of Gibbsian ensembles 
\cite{penrose1971rigorous}, 
it was soon clear that metastability 
is a genuine dynamical phenomenon which needs to be described 
by means of non--equilibrium Statistical Mechanics ideas. 
The first kinetical approach 
to metastability is the Becker--Doring theory \cite{BeckerDoring}, 
which dates back to 1935. 
However, almost half a century was needed
to arrive to the first rigorous mathematical discussion of metastable states 
in the case of 
a simple mean--field spin system \cite{cassandro1984metastable}, 
i.e., the Curie--Weiss model. The necessity of a dynamical approach 
was also pointed out by many numerical studies; we refer to the 
review \cite{RikvoldGorman} and to references therein. 

In \cite{cassandro1984metastable} the question of metastable states 
is posed in the framework of a stochastic spin system 
whose evolution is a Markov chain defined as a
reversible Glauber dynamics.
An external field is introduced in the Hamiltonian to select the 
\emph{stable} homogeneous 
phase (corresponding to the minimum of the Hamiltonian). 
As initial condition of the dynamics, the opposite state is considered,
and its metastable character is shown by proving that the time
needed by the system to hit for the first time the stable state 
is exponentially large with respect to the inverse temperature parameter. 
The method proposed in \cite{cassandro1984metastable} 
is today known as \emph{pathwise approach} and has been used to 
study metastablity in a large variety of models. 
The first paper in which it was applied with success to 
a model with a physically acceptable short range interaction is 
\cite{neves1991critical} in which the existence of metastable states
was proven for the Ising model. 

The basic idea of the pathwise approach is searching for the 
optimal path in the configuration space connecting the 
metastable state to the stable one and computing its energy height. 
The time needed by the system to perform the transition from the 
metastable to the stable state, i.e., the \emph{exit time},
is a random variable whose mean value can be estimated by an exponential
function of the energy barrier that the system must overcome during the
transition. The method provides also the portion of the 
configuration space explored by the system before  
performing the transition to the stable state 
and the tube of trajectories, called the \emph{exit path}, followed 
during the exit excursion. Along the exit path the system 
necessarily visits those particular configurations, called 
the \emph{critical configurations}, at which the optimal path 
attains its maximum. In the original version of the theory 
this was achieved via a detailed study of the configuration space 
and a suitable definition of \emph{basin of attraction} of the metastable
state based on the notion of cycles. 
The pathwise approach was further developed in 
\cite{olivieri1995markov, olivieri1996markov, scoppola1994metastability}, 
see also \cite{olivieri2005large}, and several techniques were introduced 
to simplify the application of the method. It is worth mentioning that, 
independently, a similar cycle theory was derived in 
\cite{catoni1997exit, catoni1999simulated} and applied to reversible 
Metropolis dynamics and to simulated annealing 
\cite{catoni1992parallel, trouve1996rough}. 

The general properties of the pathwise approach were further analyzed 
in \cite{manzo2004essential, cirillo2013relaxation, cirillo2015metastability, fernandez2015asymptotically, fernandez2016conditioned} 
to disentangle the study of the transition time from that of the typical 
trajectories and to treat irreversible system. 
This method has been used to study the metastable behavior of the 
Ising model with isotropic and anisotropic interaction, in different 
dimensions, with different external magnetic fields evolving 
according to Glauber dynamics, in \cite{arous1996metastability, kotecky1994shapes, kotecky1993droplet, nardi1996low, neves1991critical, neves1992behavior, olivieri2005large, jovanovski2017metastability}. 
Moreover, it has been used also for a variety of other models evolving 
according to Glauber dynamics, such as the Blume Capel model 
in \cite{cirillo1996metastability, cirillo2013relaxation}, the 
Potts model in \cite{nardi2019tunneling,bet2021potts}, and hard--core model 
in \cite{nardi2016hitting, zocca2019tunneling, den2018metastability}. 
Other applications of the pathwise approach are present 
in \cite{cirillo1998metastability, hollander2000metastability, den2003droplet, gaudilliere2005nucleation} 
for the Metropolis dynamics and in 
\cite{cirillo2008metastability, cirillo2008competitive, bet2021effect}
for parallel dynamics.

The focus of this paper is on the comparison between the metastable 
behavior of \emph{serial} or \emph{asynchronous} dynamics, i.e, stochastic 
spin systems in which the configurations are updated a site at a time, 
and \emph{parallel} or \emph{synchronous} dynamics, i.e., stochastic 
spin systems in which at each time step all the 
spins are simultaneously updated. Although the general ideas are similar, 
the parallel case is utterly more puzzling due to the very intricate 
structure of the possible paths that can be 
followed in the configuration space by the system during its random motion.
In Section~\ref{s:modelli} we consider the Hamiltonian of the 
2D Ising model and we show how it is possible to construct different 
dynamics by allowing or forbidding simultaneous spin updating.
The structure of their critical configurations and optimal paths 
are discussed in Section~\ref{s:critiche}.

Before closing the Introduction, it is worth mentioning 
two more approaches to the rigorous mathematical description of 
metastability which have been developed in the last decades.
One is known as 
the \emph{potential--theoretic approach} and is 
based on the seminal papers
\cite{mathieupicco,bovier2002metastability}. 
We refer to \cite{bovier2016metastability} for an extensive discussion  
of this method 
and of its applications to several models. In the potential--theoretic 
approach the estimate of the hitting time is achieved 
through the use of the Dirichlet form and the 
spectral properties of the transition matrix. One of the advantages 
of this method is that it provides an estimate of the expected value 
of the transition time including the prefactor, by exploiting a 
detailed knowledge of the critical configurations, 
see \cite{bovier2004metastability, bovier2016metastability}. 
This method has been applied in \cite{baldassarri2022,bashiri2017note, boviermanzo2002metastability, cirillo2017sum, bovier2006sharp, den2012metastability} to Metropolis dynamics and in \cite{nardi2012sharp} to parallel dynamics.

Finally, we mention the more recent \emph{trace method} firstly 
introduced in 
\cite{beltran2010tunneling}, that extends in some sense the 
path--wise approach and the potential theoretic approach. Differently from the pathwise approach, it  does not rely on
large deviations estimates, so that it can be used to study models where the ratio between the jump rates are
not exponential in the scaling parameter (e.g.,  condensing zero--range processes). Moreover, differently from the potential approach the
method does not depend on a reversibility assumption. The main idea is to consider a reduction of the process by removing rapid fluctuations
from the trajectory. The authors considered indeed the trace
of the process on the metastable states. 
Then, the metastability behavior is  examined through a martingale problem
that controls the convergence of the trace process. Moreover,  based on the martingale
characterization of Markov processes, a  sufficient conditions for metastability can be given (see \cite{landim2019} for a general review on the method and its applications). 

\section{Models}
\label{s:modelli} 
\par\noindent
Let $\Lambda$ be a finite square 
of $\mathbb{Z}^2$ 
with periodic boundary conditions, namely, a two--dimensional finite 
torus. With each site $i\in\Lambda$ it is associated a 
\emph{spin variable} $\sigma(i)\in\{-1,+1\}$.
We denote by $\Omega=\{-1,+1\}^\Lambda$ the \emph{configuration} or 
\emph{state} space and we call \emph{configuration} or \emph{state}
any element $\sigma\in\Omega$.
We say that two sites are \emph{nearest neighbors} if and only if their 
Euclidean distance is equal to one. 
Given $i\in\Lambda$, 
we consider 
the \emph{shift operator} 
$\Theta_i:\Omega\to\Omega$ 
which shifts a configuration so that the site $i$ is mapped to the origin $0$, 
that is to say 
$(\Theta_i(\sigma))_j=\sigma_{i+j}$ for any $j\in\Lambda$.
Given $\Delta\subset\Lambda$ we denote by $\sigma_\Delta$ the 
restriction to $\Delta$ of a configuration $\sigma\in\Omega$.
Given $\sigma\in\Omega$
we denote by $\sigma^s$, for $s\in\{-1,+1\}$, the configuration obtained 
by setting to $s$ the value of the spin at the origin, namely, 
$\sigma^s(0)=s$ and $\sigma^s(i)=\sigma(i)$ for any 
$i\in\Lambda\setminus\{0\}$. 

A \emph{pairwise interaction} is a collection of real numbers 
$J_{ij}$, for any $i,j\in\Lambda$, for $i\neq j$,
symmetrical and translationally invariant,
which means that 
$J_{ij}=J_{ji}$ and
$J_{ij} = J_{i+k,j+k}$
for all $k\in\mathbb{Z}^2$.
We shall also assume that the interaction is \emph{finite range}, 
i.e., there exists $I\subset\Lambda\setminus\{0\}$ (not depending on 
the size of $\Lambda$) such that 
$J_{0i}\neq0$ for any $i\in I$ and 
$J_{0i}=0$ otherwise: it is worth noting that the symmetry of the interaction
implies that $I$ is symmetric with respect to the origin. 
We thus define the Hamiltonian 
\begin{equation}
\label{mod000}
H(\sigma)
=
-\frac{1}{2}\sum_{\newatop{i,j\in\Lambda:}{i\neq j}}J_{ij}\sigma(i)\sigma(j)
-h\sum_{i\in\Lambda}\sigma(i)
,
\end{equation}
with $h\in\mathbb{R}$ the \emph{external magnetic field},
and consider on the single spin space the probability 
distribution 
\begin{equation}
\label{mod010}
f_{T,\sigma}(s)
=
\frac{\exp\{-H(\sigma^s)/T\}}
  {\exp\{-H(\sigma^s)/T\}
   +\exp\{-H(\sigma^{-s})/T\}}
\end{equation}
for any $\sigma\in\Omega$.
Note that the probability distribution $f_{T,\sigma}$
can be rewritten as
\begin{equation}
\label{mod020}
f_{T,\sigma}(s)
=
\frac{1}
  {1+\exp\{[H(\sigma^s)-H(\sigma^{-s})]/T\}}
=
\frac{1}
  {1+\exp\{-2s(\sum_{i\in I}J_{0i}\sigma(i)+h)/T\}}
,
\end{equation}
which then implies
\begin{equation}
\label{mod030}
f_{T,\sigma}(s)
=
\frac{1}{2}
\Big\{1+s\tanh\Big[\frac{1}{T}
   \Big(\sum_{i\in I}J_{0i}\sigma(i)+h\Big)\Big]\Big\}
.
\end{equation}

We shall consider stochastic evolutions in $\Omega$ modelled
as Markov chains in which single spins are updated according to the 
probability distribution \eqref{mod010}. 
The dynamics can be implemented by allowing one 
spin change at a time or the simultaneous change of all the spins 
on the lattice, the dynamics will be respectively 
called \emph{asynchronous} or \emph{synchronous}.

\subsection{Asynchronous dynamics}
\label{s:asincrono} 
\par\noindent
The \emph{heat bath} dynamics is defined as the discrete time 
Markov chain $\sigma_t\in\Omega$, with $t\in\mathbb{Z}_+$, 
with updating rule defined as follows: at time $t$ a site $i$
is chosen at random with uniform probability $1/|\Lambda|$ and 
the configuration $\sigma_t$ is constructed by letting 
$\sigma_t(j)=\sigma_{t-1}(j)$ for any $j\neq i$ and 
$\sigma_t(i)=s$ with probability $f_{T,\Theta_i\sigma_{t-1}}(s)$.
In other words, the heat bath dynamics is the discrete time 
Markov chain 
with transition probability defined as follows:
for $\sigma,\eta\in\Omega$ such that $\sigma\neq\eta$
\begin{equation}
\label{sin000}
p_T(\sigma,\eta)=
\left\{
\begin{array}{ll}
\frac{1}{|\Lambda|}f_{T,\Theta_i\sigma}(\eta(i))
&
\textup{if }
\exists i\in\Lambda:
\eta(i)\neq\sigma(i)
\textup{ and }
\eta(j)=\sigma(j)
\textup{ for }
j\neq i
\\
0
&
\textup{otherwise}
\end{array}
\right.
\end{equation}
and 
\begin{equation}
\label{sin010}
p_T(\sigma,\sigma)=
1-\sum_{\eta\in\Omega\setminus\{\sigma\}}p_T(\sigma,\eta).
\end{equation}

As it is well known \cite[Section~4.5.1]{newmanbarkema}, the heat bath dynamics is \emph{reversible}
with respect to the \emph{Gibbs measure} on $\Omega$ 
\begin{equation}
\label{sin020}
\mu_T(\sigma)
=
\frac{1}{Z_T}e^{-H(\sigma)/T}
\end{equation}
where the \emph{partition function} is
\begin{equation}
\label{sin030}
Z_T=
\sum_{\sigma\in\Omega} e^{-H(\sigma)/T}
,
\end{equation}
that is to say, the \emph{detailed balance condition} 
\begin{equation}
\label{sin040}
\mu_T(\sigma) p_T(\sigma,\eta)=\mu_T(\eta)p_T(\eta,\sigma)
\end{equation}
is satisfied 
for any $\sigma,\eta\in\Omega$. 
This, together with the fact that the Markov chain is irreducible 
and the state space is finite, implies that the stationary measure 
is unique and it is given by the Gibbs measure. 

Asynchronous models are typically used in Statistical Mechanics 
to introduce stochastic versions of statistical spin systems in order 
to study how equilibrium is approached. Often, 
for efficiency reasons, the Metropolis algorithm 
\cite[Section~3.1]{newmanbarkema} is preferred to heat bath 
and to other similar rules.

A crucial role in the study of the metastable behavior of the 
dynamics is played by the so called energy cost. Consider two 
configurations, $\sigma$ and $\eta$, differing by only the spin at site $i$;
we have that, in the limit $T\to0$,
$p(\sigma,\eta)\approx1/|\Lambda|$ if $H(\eta)<H(\sigma)$
and 
$p(\sigma,\eta)\approx\exp\{-[H(\eta)-H(\sigma)]/T\}/|\Lambda|$ 
if $H(\eta)>H(\sigma)$. Then it is reasonable to define 
the \emph{energy cost} of the transition from $\sigma$ to 
$\eta$ as the quantity
$\Delta(\sigma,\eta)=0$ if $H(\eta)<H(\sigma)$
and 
$\Delta(\sigma,\eta)=H(\eta)-H(\sigma)$ if $H(\eta)>H(\sigma)$.

\subsection{Synchronous dynamics}
\label{s:sincrono} 
\par\noindent
A model in which all the spins are updated at each time independently and 
simultaneously with the probability distribution \eqref{mod010}
can be defined as the
Markov chain $\sigma_t\in\Omega$, with $t\in\mathbb{Z}_+$, 
with transition matrix
\begin{equation}
\label{asi000}
p_T(\sigma,\eta)=
\prod_{i\in\Lambda}
f_{T,\Theta_i\sigma}(\eta(i))
.
\end{equation}
This model is an example of \emph{reversible Probabilistic 
Cellular Automata} (PCA), see, e.g., 
\cite{daipraetal,derrida,lebowitzmaesspeer}. 
We mention that the class of reversible PCA is slightly larger, indeed, 
the constraint that the set $I$ in \eqref{mod030} 
does not contain the origin can be relaxed; when the origin is 
considered in $I$, 
as, for instance, in \cite{bigelis1999critical,cirillo2008metastability} its 
contribution is called \emph{self--interaction} term. It does not appear in 
our derivation of reversible PCA, since we started from the Hamiltonian 
\eqref{mod000} where it would appear simply as a constant additive 
irrelevant contribution \cite[Section~4.5.1, equation (4.38)]{newmanbarkema}.

As it is proven 
in \cite{grinsteinetal,kozlovvasiljev}, reversible Probabilistic 
Cellular Automata satisfy the 
detailed balance condition 
\begin{equation}
\label{asi010}
\mu_T(\sigma) p_T(\sigma,\eta)=\mu_T(\eta)p_T(\eta,\sigma)
\end{equation}
with respect to the Gibbs measure on $\Omega$ 
\begin{equation}
\label{asi020}
\mu_T(\sigma)
=
\frac{1}{Z_T}e^{-G_T(\sigma)/T}
\end{equation}
where 
\begin{equation}
\label{asi030}
Z_T=
\sum_{\sigma\in\Omega} e^{-G_T(\sigma)/T}
\end{equation}
is the partition function
and 
\begin{equation}
\label{asi040}
G_T(\sigma)
=-h\sum_{i\in\Lambda}\sigma(i)
	   -T\sum_{i\in\Lambda}
	     \log\cosh\Big[\frac{1}{T}
                \Big(\sum_{j\in i+I}J_{ij}\sigma(j)+h\Big)\Big]
\end{equation}
for any $\sigma\in\Omega$.
Thus, even in the synchronous case we have that 
the stationary measure 
is unique, but it is different from the Gibbs measure \eqref{sin020} found 
for the Statistical Mechanics model considered in the 
asynchronous case. 

It is interesting to note that the Gibbs measure \eqref{asi020} 
is such that $G_T=-T\log(Z_T\mu_T)$ depends on the temperature,
which is not the case in standard 
Statistical Mechanics models. 
On the other hand, the global minima of the function 
\begin{equation}
\label{asi050}
H(\sigma)
=
\lim_{T\to0}G_T(\sigma)
=-h\sum_{i\in\Lambda}\sigma(i)
	   -\sum_{i\in\Lambda}
                \Big|
                \sum_{j\in i+I}J_{ij}\sigma(j)+h
                \Big|
\end{equation}
configurations 
are the configurations in which the dynamics will be trapped at low 
temperature.
By abusing the notation, 
the function $H$ will be called \emph{Hamiltonian} or \emph{energy} and its 
global minima will be called \emph{ground states}.
Moreover, a crucial role in our discussion will be played 
by the \emph{cost function}
\begin{equation}
\label{asi060}
\Delta(\sigma,\eta)
=
-\lim_{T\to0}T\log\pi_T(\sigma,\eta)
=
\sum_{\newatop{i\in\Lambda:}{\eta(i)[\sum_{j\in i+I}J_{ij}\sigma(j)+h]<0}}
2\Big|\sum_{j\in i+I}J_{ij}\sigma(j)+h\Big|
,
\end{equation}
since, as it as been proven in \cite[Section~2.6]{cirillo2008metastability},
for the transition matrix it is possible to prove the so--called 
Friedlin--Wentzel condition 
\begin{equation}
\label{asi070}
e^{-\Delta(\sigma,\eta)/T-\gamma(T)/T}
\le
\pi_T(\sigma,\eta)
\le
e^{-\Delta(\sigma,\eta)/T+\gamma(T)/T}
\end{equation}
with $\gamma(T)\to0$ as $T\to0$.
Equation \eqref{asi070}, together with the equality
\begin{equation}
\label{asi080}
H(\sigma)+\Delta(\sigma,\eta)=
H(\eta)+\Delta(\eta,\sigma), 
\end{equation}
which follows immediately from \eqref{asi010}, \eqref{asi020},
\eqref{asi050}, \eqref{asi060}, 
allows us to interpret $\Delta(\sigma,\eta)$ as the energy cost 
that the chain has to pay in the jump from $\sigma$ to $\eta$. 
We note that it is possible that both $\Delta(\sigma,\eta)$
and $\Delta(\eta,\sigma)$ are positive. This is not the case in 
the asynchronous case, in which only one of the two can be positive. 

\subsection{The Ising case}
\label{s:ising} 
\par\noindent
In this paper we shall focus our discussion to the 
Ising case, namely, we assume that the interaction
is $1$ for nearest neighbors spins and $0$ otherwise, 
which mean that the set $I$ is equal to set of the four
neighboring sites of the origin.
With this choice the Hamiltonian \eqref{mod000} of the asynchronous model
is the standard 2D Ising Hamiltonian
\begin{equation}
\label{isi000}
H(\sigma)
=
-\frac{1}{2}\sum_{\newatop{i,j\in\Lambda:}{|i-j|=1}}\sigma(i)\sigma(j)
-h\sum_{i\in\Lambda}\sigma(i)
.
\end{equation}

On the other hand, 
the energy of the 
synchronous version of the dynamics reads
\begin{equation}
\label{isi005}
H(\sigma)
=-h\sum_{i\in\Lambda}\sigma(i)
	   -\sum_{i\in\Lambda}
                \Big|
                \sum_{\newatop{j\in\Lambda:}{|j-i|=1}}
                \sigma(j)+h
                \Big|
.
\end{equation}
In Statistical Mechanics models, Hamiltonians are usually written in 
terms of coupling constants. Following \cite{cirilloetal2014}, see 
also \cite{hallerkennedy,cirillostramaglia}, 
for the Hamiltonian
\eqref{isi005}
we get 

\begin{equation}
\label{isi010}
\begin{array}{ll} 
{\displaystyle{ 
H(\sigma)=}} 
& 
{\displaystyle{ 
 \!\!\!
-J_{.} 
\sum_{x\in\Lambda}\sigma(x) 
-J_{_{\langle\langle\rangle\rangle}} 
\sum_{\langle\langle x y\rangle\rangle}\sigma(x)\sigma(y) 
-J_{_{\langle\langle\langle \rangle\rangle\rangle}} 
\sum_{\langle\langle\langle x y\rangle\rangle\rangle}\sigma(x)\sigma(y) 
}}\\ 
& 
{\displaystyle{ 
 \!\!\!
-J_{_{\triangle}} 
\sum_{\triangle_{xyz}}\sigma(x)\sigma(y)\sigma(z) 
-J_{_{\diamondsuit}} 
\sum_{\diamondsuit_{xywz}}\sigma(x)\sigma(y)\sigma(w)\sigma(z)  
}} 
\\ 
\end{array} 
\end{equation}
where the coupling constants
\begin{equation} 
\label{isi020}
  J_{.} = \frac{5}{2} h,\;
  J_{_{\langle\langle\rangle\rangle}}= 1-\frac{1}{4}h,\;
  J_{_{\langle\langle\langle \rangle\rangle\rangle}}
      =\frac{1}{2}-\frac{1}{8}h,\;
  J_{_{\triangle}} = -\frac{1}{8}h 
  J_{_{\diamondsuit}} = -\frac{1}{2}+\frac{3}{8} h,\;
\end{equation} 
refer, respectively, to single site, next to the nearest neighbor pairs, 
third neighbor pairs, triangle, and diamond clusters (see, 
figure~\ref{f:fig02}). Note that the diamond cluster coupling is negative
for small magnetic field, this yields an anti--ferromagnetic behavior 
of the interaction that will give rise to very peculiar 
phenomena that will be discussed in the following sections. 
The triangle cluster coupling is negative, as well,
but it becomes negligible for small $h$. 

\setlength{\unitlength}{1.pt}
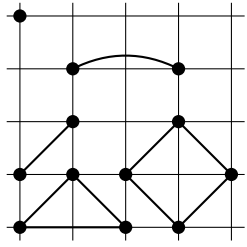
\begin{figure}
\begin{picture}(200,80)(-200,0)
\thinlines
\multiput(-20,0)(20,0){5}{\line(0,1){90}}
\multiput(-25,5)(0,20){5}{\line(1,0){90}}
\thicklines
\put(-20,25){\circle*{5}}
\put(0,45){\circle*{5}}
\put(-20,25){\line(1,1){20}}
\put(0,65){\circle*{5}}
\put(40,65){\circle*{5}}
\qbezier(0,65)(20,75)(40,65)
\put(20,25){\circle*{5}}
\put(40,45){\circle*{5}}
\put(40,5){\circle*{5}}
\put(60,25){\circle*{5}}
\put(20,25){\line(1,1){20}}
\put(20,25){\line(1,-1){20}}
\put(60,25){\line(-1,1){20}}
\put(60,25){\line(-1,-1){20}}
\put(-20,85){\circle*{5}}
\put(-20,5){\circle*{5}}
\put(0,25){\circle*{5}}
\put(20,5){\circle*{5}}
\put(-20,5){\line(1,1){20}}
\put(-20,5){\line(1,0){40}}
\put(0,25){\line(1,-1){20}}
\end{picture}
\caption{Schematical representation of coupling constants \eqref{isi020}.}
\label{f:fig02}
\end{figure}

We stress that implementing the dynamics in asynchronous or synchronous 
ways yield completely different stochastic models. 
We have already noticed that the Hamiltonians are different, 
the simple 2D Ising Hamiltonian has to be compared to the 
complicated Hamiltonian \eqref{isi010} found in the synchronous case. 
Even the cost functions are pretty different: in the asynchronous case 
it reduces to the positive part of the energy difference between 
configurations 
differing for the value of one spin, whereas in the synchronous case 
it is defined for any pair of configurations and it is given by the 
baroque formula \eqref{asi060}.
But we want to remark that in the latter case the energy cost 
is defined for any pair of configurations because in a single 
step the system can jump from any configuration to any other.
This is the essential difficulty of studying metastability 
for synchronous dynamics, the fact that the structure of 
the trajectories followed by the system in the configuration space 
is utterly complicated due to the fact that any transition is possible 
in a single time step. 

\section{Metastable state and critical configurations}
\label{s:critiche} 
\par\noindent
As already mentioned in Section~\ref{s:intro} 
the first rigorous and full description of metastable behavior 
from the mathematical point of view
dates back to the paper \cite{cassandro1984metastable}
where the theory of the pathwise approach was firstly introduced. 
Other widely applied methods are nowadays known as 
the potential theoretic approach and the trace method, 
respectively proposed in \cite{mathieupicco,bovier2002metastability}
and \cite{beltran2010tunneling}. For a comprehensive 
discussion of the pathwise and the potential theoretic 
approach we refer the reader to the books 
\cite{olivieri2005large,bovier2016metastability}.

Here, we shall review some metastability results adopting the 
pathwise point of view, in particular we shall follow the strategy
refined in \cite{manzo2004essential} for the synchronous dynamics 
with Metropolis updating algorithm and extended in 
\cite{cirillo2015metastability} to a much larger class of
Markov chains, including not reversible dynamics. 
This theory applies both to 
the heat bath Glauber dynamics and to the reversible PCA.
We warn the reader that we use the same notation ($H$ for Hamiltonian, 
$\Delta$ for the energy cost, etc.) for the two cases, but, 
depending on the model one has in mind, the correct related quantities should 
be used. 

\setlength{\unitlength}{0.8pt}
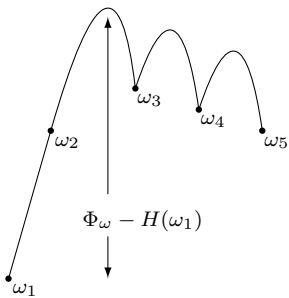
\begin{figure}
\begin{picture}(200,120)(-100,0)
 \qbezier(130,10)(140,45)(150,80)
 \put(130,10){\circle*{3}}
 \put(132,3){\scriptsize{$\omega_1$}}
 \qbezier(150,80)(180,185)(190,100)
 \put(150,80){\circle*{3}}
 \put(152,73){\scriptsize{$\omega_2$}}
 \qbezier(190,100)(210,160)(220,90)
 \put(190,100){\circle*{3}}
 \put(191,93){\scriptsize{$\omega_3$}}
 \qbezier(220,90)(240,150)(250,80)
 \put(220,90){\circle*{3}}
 \put(221,83){\scriptsize{$\omega_4$}}
 \put(250,80){\circle*{3}}
 \put(251,73){\scriptsize{$\omega_5$}}
 \put(177,50){\vector(0,1){84}}
 \put(165,35){\scriptsize{$\Phi_\omega-H(\omega_1)$}}
 \put(177,30){\vector(0,-1){20}}
 \end{picture}
\caption{Schematical representation of the height of a path.}
\label{f:fig03}
\end{figure}

\subsection{General results}
\label{s:generale} 
\par\noindent
A \emph{path of length} $n\ge1$ 
is a sequence $\{\omega_1,\dots\omega_n\}\in\Omega^n$ such 
that $p_T(\omega_i,\omega_{i+1})>0$ for any $i=1,\dots,n-1$. 
We denote by $\mathcal{F}$ the set of all loop free paths and, 
given $A,A'\subset\Omega$ not empty, we let 
$\mathcal{F}(A,A')$ be the set of all loop free paths with 
first configuration in $A$ and last configuration in $A'$.
Given a path $\omega$ of length $n$, we define the
\emph{height} of the path as
\begin{equation}
\label{gen000}
\Phi_\omega=\max_{i=1,\dots,n-1}[H(\omega_i)+\Delta(\omega_i,\omega_{i+1})]
.
\end{equation}
Given $A,A'\subset\Omega$, the 
\emph{communication height} $\Phi(A,A')$ between $A$ and $A'$ 
is defined as 
\begin{equation}
\label{gen020}
\Phi(A,A')=\min_{\omega\in\mathcal{F}(A,A')}\Phi_\omega
\end{equation}

We define the \emph{stability level} $V_\sigma$ of any configuration 
$\sigma\in\Omega$ as the minimal height, with respect to 
$H(\sigma)$, that must be overcome 
by paths connecting $\sigma$ to the set of configurations at energy 
smaller than $H(\sigma)$, namely, 
\begin{equation}
\label{gen030}
V_\sigma=\Phi(\{\sigma\},\{\eta\in\Omega:\,H(\eta)<H(\sigma)\})-H(\sigma)
.
\end{equation}

We are now ready to give the key notion of metastable state.
We let $\Omega^{\textup{s}}$ be the set of the absolute minima of the 
Hamiltonian, namely, the set of ground states, 
and 
define the \emph{maximal stability level}
\begin{equation}
\label{gen040}
\Gamma=\max_{\sigma\in\Omega\setminus\Omega^{\textup{s}}}V_\sigma>0
\end{equation}
and 
the set of \emph{metastable} states 
\begin{equation}
\label{gen050}
\Omega^{\textup{m}}=\{\sigma\in\Omega\setminus\Omega^{\textrm{s}}:
                                                   \,V_\sigma=\Gamma\}.
\end{equation}

The set $\Omega^{\textup{m}}$ deserves the name of set of metastable 
states since it is possible to prove the following theorem
\cite[Theorem~2.1]{cirillo2015metastability}: for any 
$\sigma\in\Omega^{\textup{m}}$, for any $\varepsilon>0$ we have that 
\begin{equation}
\label{gen060}
   \lim_{T\to0}
     \mathbb{P}_{\sigma}
       (
        e^{(\Gamma-\varepsilon)/T}
        <\tau_{\Omega^{\textup{s}}}<
        e^{(\Gamma+\varepsilon)/T}
       )
   =1
,
\end{equation}
where $\mathbb{P}_\sigma$ is the probability for the chain $\sigma_t$
started at $\sigma$ and 
the random variable $\tau_{\Omega^{\textup{s}}}$ is the 
\emph{first hitting time} to $\Omega^{\textup{s}}$ for the dynamics 
started at $\sigma$, i.e., 
$\tau_{\Omega^{\textup{s}}}=\inf\{t\ge0:\,\sigma_t\in\Omega^{\textup{s}}\}$.
In words, equation \eqref{gen060} states that the time needed by 
the system to exit the metastable state and touch the ground state 
is, controlled in probability, of order $\exp\{\Gamma/T\}$. 
Thus, \eqref{gen060} gives a mathematically rigorous meaning 
to the statement ``the exit time from the metastable state 
is of order $\exp\{\Gamma/T\}$".

The pathwise approach provides also an estimate for the mean 
value $\mathbb{E}_\sigma[\tau_{\Omega^{\textup{s}}}]$, 
indeed, it is possible to prove that 
\begin{equation}
\label{gen070}
   \lim_{T\to0}
     T\log \mathbb{E}_\sigma[\tau_{\Omega^{\textup{s}}}]
=\Gamma, 
\end{equation}
see
\cite[Theorem~2.2]{cirillo2015metastability}.

The exit time is, for sure, the main property of metastable states 
that one want to compute. Another relevant property concerns 
the path followed by the system to exit the metastable state.
Think to a super--saturated vapor: how does it perform the transition
to the liquid stable phase? Will it happen through the coalescence 
of small droplets of liquid phase appeared throughout the whole 
volume occupied by the system? Or will it happen through a sudden formation 
of a sufficiently large droplet?
These questions can be answered in the framework of the pathwise 
approach, indeed it is possible to identify configurations, 
called \emph{critical},
that must be necessarily visited during the excursion from the metastable
to the stable state. These special configurations will give a clear 
indication of the mechanism of the transition from the metastable to the 
stable state. 

To make this formal, 
given $A,A'\subset\Omega$ not empty, we define 
the set of \emph{optimal paths} connecting $A$ to $A'$, and denote it by 
$\mathcal{F}_\textup{o}(A,A')$, as the set of loop free paths 
$\omega\in\mathcal{F}(A,A')$ such that $\Phi_\omega=\Phi(A,A')$.
That is to say, an optimal path connecting $A$ to $A'$ is a path
starting in $A$, ending in $A'$, and having maximal height equal to 
the communication height between $A$ and $A'$. 

Morally, the critical configurations are those configurations 
where the optimal paths attain the maximal height. Unfortunately, 
as remarked above and depicted in figure~\ref{f:fig03}, the maximal
height of a path in general does not correspond to the energy 
of one of the configurations forming the path. This is true in the 
case of Glauber dynamics \cite{manzo2004essential}, but it is not 
necessarily true in a 
more general set--up including the PCA case. Thus, a more sophisticated 
notion of critical configuration is needed
\cite{cirillo2015metastability}: given 
$\eta\in\Omega$ 
and 
$A'\subset\Omega$ 
the set of \emph{saddles} $\mathcal{S}(\{\eta\},A')$ between
$\eta$ and $A'$ is the set of configurations $\xi$ such that 
there exists an optimal path 
$\omega\in\mathcal{F}_\textup{o}(\{\eta\},A')$ 
and a configuration $\zeta$ such that $\xi$ follows $\zeta$ in the 
path $\omega$ and $H(\zeta)+\Delta(\zeta,\xi)=\Phi(\{\eta\},A')$.

Among all the possible saddles, a relevant role is played by those 
that must be necessarily visited by optimal paths: given 
$\eta\in\Omega$ 
and 
$A'\subset\Omega$ 
a subset $W\subset\mathcal{S}(\{\eta\},A')$ is a \emph{gate} 
for $\eta$ and $A'$ if and only if every optimal path 
in $\mathcal{F}_\textup{o}(\{\eta\},A')$ intersects $W$.
Moreover, a gate $W$ is \emph{minimal} if and only if for any 
$W'\subset W$ and $W'\neq W$ there exists an optimal path 
which does not intersect $W'$.
The fact that gates must be necessarily visited during the 
transition from the metastable to the stable state is proven in 
\cite[Theorem~2.4]{cirillo2015metastability} stating that 
given $\sigma\in\Omega^\textup{m}$ and 
$W$ a minimal gate for $\sigma$ and $\Omega^\textup{s}$, we have that 
there exists $c>0$ such that
\begin{equation}
\label{gen080}
\mathbb{P}_\sigma[\tau_{W}>\tau_{\Omega^\textup{s}}]
\le e^{-c/T}
,
\end{equation}
where $\tau_W$ is the first hitting time to $W$ for the dynamics started 
at $\sigma$.

We mention that in the framework of the pathwise approach it is 
also possible to characterize the behavior of the system before 
it performs the transition to the metastable state. Indeed, 
by using the theory of cycles, it is possible to define 
the basin of attraction of the metastable state and to prove that 
in a time smaller than the exit time the system is confined to move 
within such a basin of attraction. 

Finally, we remark that the estimate provided by the pathwise approach 
for the exit time is rather rough, in the sense that it is 
given at the level of logarithmic equivalence. A more refined 
estimate can be provided in the framework of the potential theoretic 
approach and using the trace method. The typical result that can be proven
is the following: for $T\to0$, 
\begin{equation}
\label{gen090}
\frac{\mathbb{E}_T[\tau_{\Omega^\textup{s}}]}{\exp\{\Gamma/T\}} = 
C(\Lambda)[1+o(1)]
\end{equation}
where $C(\Lambda)$ is a constant, depending on the volume $\Lambda$, 
which can be computed in terms of the number of saddles 
between the starting metastable state $\sigma$ and 
the set of ground states $\Omega^\textup{s}$.

The results that we have discussed in in this section
are general, but the
model dependent inputs that will be presented in Sections~\ref{s:met-ising} 
and \ref{s:met-pca} are necessary if one wants 
to describe the metastable behavior of a specific system.
This is precisely the idea on which
the paper \cite{manzo2004essential} is based 
and which was further developed in \cite{cirillo2015metastability}: 
disentangling the proof of the general properties of metastable 
states from that of the model dependent inputs. 
These model dependent inputs, necessary to achieve the full 
characterization of the metastable behavior of a particular model,
are the set of metastable states and the optimal paths 
connecting such states to the set of stables one. Once we get 
this, all the properties follow from the general theory. 
 
\subsection{Metastable behavior of the 2D Ising model}
\label{s:met-ising} 
\par\noindent
We consider the asynchronous dynamics (heat bath) introduced 
in Section~\ref{s:asincrono} for the standard 2D Ising Hamiltonian 
considered in Section~\ref{s:ising} with $0<h\ll 1$ such that 
$2/h$ is not integer. 
This assumption reduces the number
of degenerate critical configurations; we refer to \cite{manzo2004essential}
for a thorough discussion of the case in which $2/h$ is integer and
to \cite{neves1991critical}, page 213,
where some comments on this singular
point are reported.
Although 
the metastable behavior for 
such a model was first studied in \cite{neves1991critical}
in the case of the Metropolis single site updating rule, we describe 
here the main results using the language developed in the 
previous sections. 

\setlength{\unitlength}{1.2pt}
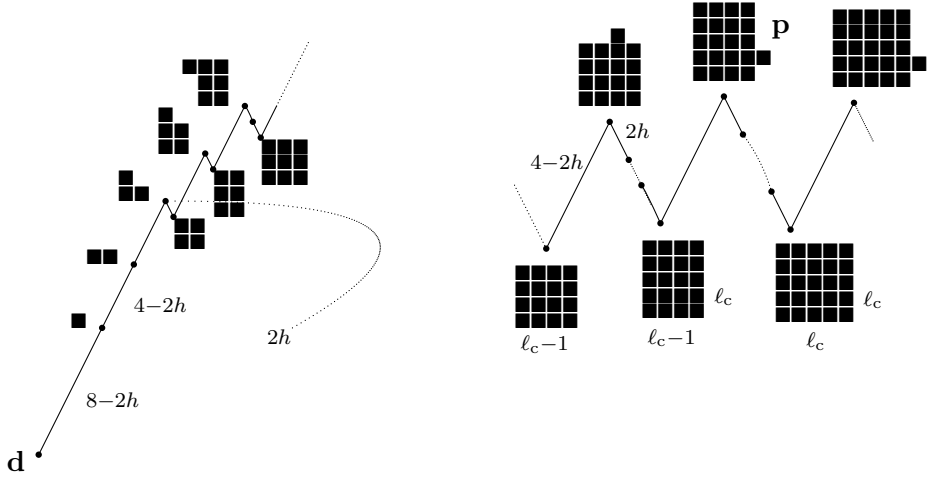
\begin{figure}
 \begin{picture}(300,160)(-50,-10)
 \thinlines
 \put(-10,-5){${\textstyle \textbf{d}}$}
 \put(0,0){\circle*{2}}
 \put(0,0){\line(1,2){20}}
 \put(20,40){\circle*{2}}
 \put(10,40){${\scriptstyle \blacksquare}$}
 \put(15,15){${\scriptstyle 8-2h}$}
 \put(20,40){\line(1,2){10}}
 \put(30,60){\circle*{2}}
 \put(20,60){${\scriptstyle \blacksquare}$}
 \put(15,60){${\scriptstyle \blacksquare}$}
 \put(30,45){${\scriptstyle 4-2h}$}
 \put(30,60){\line(1,2){10}}
 \put(40,80){\circle*{2}}
 \put(30,80){${\scriptstyle \blacksquare}$}
 \put(25,80){${\scriptstyle \blacksquare}$}
 \put(25,85){${\scriptstyle \blacksquare}$}
 \qbezier(40,80)(41,78)(42.5,75)
 \put(42.5,75){\circle*{2}}
 \put(42.5,70){${\scriptstyle \blacksquare}$}
 \put(47.5,70){${\scriptstyle \blacksquare}$}
 \put(42.5,65){${\scriptstyle \blacksquare}$}
 \put(47.5,65){${\scriptstyle \blacksquare}$}
 \put(42.5,75){\line(1,2){10}}
 \put(52.5,95){\circle*{2}}
 \put(42.5,95){${\scriptstyle \blacksquare}$}
 \put(37.5,95){${\scriptstyle \blacksquare}$}
 \put(42.5,100){${\scriptstyle \blacksquare}$}
 \put(37.5,100){${\scriptstyle \blacksquare}$}
 \put(37.5,105){${\scriptstyle \blacksquare}$}
 \qbezier(52.5,95)(53.5,93)(55,90)
 \put(55,90){\circle*{2}}
 \put(55,85){${\scriptstyle \blacksquare}$}
 \put(60,85){${\scriptstyle \blacksquare}$}
 \put(55,80){${\scriptstyle \blacksquare}$}
 \put(60,80){${\scriptstyle \blacksquare}$}
 \put(55,75){${\scriptstyle \blacksquare}$}
 \put(60,75){${\scriptstyle \blacksquare}$}
 \put(55,90){\line(1,2){10}}
 \put(65,110){\circle*{2}}
 \put(55,110){${\scriptstyle \blacksquare}$}
 \put(50,110){${\scriptstyle \blacksquare}$}
 \put(55,115){${\scriptstyle \blacksquare}$}
 \put(50,115){${\scriptstyle \blacksquare}$}
 \put(55,120){${\scriptstyle \blacksquare}$}
 \put(50,120){${\scriptstyle \blacksquare}$}
 \put(45,120){${\scriptstyle \blacksquare}$}
 \qbezier(65,110)(66,109)(67.5,105)
 \put(67.5,105){\circle*{2}}
 \qbezier(67.5,105)(68.5,103)(70,100)
 \put(70,100){\circle*{2}}
 \put(70,95){${\scriptstyle \blacksquare}$}
 \put(75,95){${\scriptstyle \blacksquare}$}
 \put(80,95){${\scriptstyle \blacksquare}$}
 \put(70,90){${\scriptstyle \blacksquare}$}
 \put(75,90){${\scriptstyle \blacksquare}$}
 \put(80,90){${\scriptstyle \blacksquare}$}
 \put(70,85){${\scriptstyle \blacksquare}$}
 \put(75,85){${\scriptstyle \blacksquare}$}
 \put(80,85){${\scriptstyle \blacksquare}$}
 \put(70,100){\line(1,2){5}}
 \qbezier[20](75,110)(80,120)(85,130)
 \qbezier[100](43,80)(150,80)(80,40)
 \put(72,35){${\scriptstyle 2h}$}
 \qbezier[20](150,85)(155,75)(160,65)
 \put(160,65){\circle*{2}}
 \put(150,40){${\scriptstyle \blacksquare}$}
 \put(155,40){${\scriptstyle \blacksquare}$}
 \put(160,40){${\scriptstyle \blacksquare}$}
 \put(165,40){${\scriptstyle \blacksquare}$}
 \put(150,45){${\scriptstyle \blacksquare}$}
 \put(155,45){${\scriptstyle \blacksquare}$}
 \put(160,45){${\scriptstyle \blacksquare}$}
 \put(165,45){${\scriptstyle \blacksquare}$}
 \put(150,50){${\scriptstyle \blacksquare}$}
 \put(155,50){${\scriptstyle \blacksquare}$}
 \put(160,50){${\scriptstyle \blacksquare}$}
 \put(165,50){${\scriptstyle \blacksquare}$}
 \put(150,55){${\scriptstyle \blacksquare}$}
 \put(155,55){${\scriptstyle \blacksquare}$}
 \put(160,55){${\scriptstyle \blacksquare}$}
 \put(165,55){${\scriptstyle \blacksquare}$}
 \put(152,33){${\scriptstyle \ell_{\textrm{c}}-1}$}
 \put(160,65){\line(1,2){20}}
 \put(180,105){\circle*{2}}
 \put(170,110){${\scriptstyle \blacksquare}$}
 \put(175,110){${\scriptstyle \blacksquare}$}
 \put(180,110){${\scriptstyle \blacksquare}$}
 \put(185,110){${\scriptstyle \blacksquare}$}
 \put(170,115){${\scriptstyle \blacksquare}$}
 \put(175,115){${\scriptstyle \blacksquare}$}
 \put(180,115){${\scriptstyle \blacksquare}$}
 \put(185,115){${\scriptstyle \blacksquare}$}
 \put(170,120){${\scriptstyle \blacksquare}$}
 \put(175,120){${\scriptstyle \blacksquare}$}
 \put(180,120){${\scriptstyle \blacksquare}$}
 \put(185,120){${\scriptstyle \blacksquare}$}
 \put(170,125){${\scriptstyle \blacksquare}$}
 \put(175,125){${\scriptstyle \blacksquare}$}
 \put(180,125){${\scriptstyle \blacksquare}$}
 \put(185,125){${\scriptstyle \blacksquare}$}
 \put(180,130){${\scriptstyle \blacksquare}$}
 \qbezier(180,105)(181,103)(186,93)
 \put(186,93){\circle*{2}}
 \qbezier[20](186,93)(189.5,86)(193,79)
 \put(190,85){\circle*{2}}
 \qbezier(190,85)(191,84)(196,73)
 \put(196,73){\circle*{2}}
 \put(190,43){${\scriptstyle \blacksquare}$}
 \put(195,43){${\scriptstyle \blacksquare}$}
 \put(200,43){${\scriptstyle \blacksquare}$}
 \put(205,43){${\scriptstyle \blacksquare}$}
 \put(190,48){${\scriptstyle \blacksquare}$}
 \put(195,48){${\scriptstyle \blacksquare}$}
 \put(200,48){${\scriptstyle \blacksquare}$}
 \put(205,48){${\scriptstyle \blacksquare}$}
 \put(190,53){${\scriptstyle \blacksquare}$}
 \put(195,53){${\scriptstyle \blacksquare}$}
 \put(200,53){${\scriptstyle \blacksquare}$}
 \put(205,53){${\scriptstyle \blacksquare}$}
 \put(190,58){${\scriptstyle \blacksquare}$}
 \put(195,58){${\scriptstyle \blacksquare}$}
 \put(200,58){${\scriptstyle \blacksquare}$}
 \put(205,58){${\scriptstyle \blacksquare}$}
 \put(190,63){${\scriptstyle \blacksquare}$}
 \put(195,63){${\scriptstyle \blacksquare}$}
 \put(200,63){${\scriptstyle \blacksquare}$}
 \put(205,63){${\scriptstyle \blacksquare}$}
 \put(213,48){${\scriptstyle \ell_{\textrm{c}}}$}
 \put(192,35){${\scriptstyle \ell_{\textrm{c}}-1}$}
 \put(196,73){\line(1,2){20}}
 \put(216,113){\circle*{2}}
 \put(206,118){${\scriptstyle \blacksquare}$}
 \put(211,118){${\scriptstyle \blacksquare}$}
 \put(216,118){${\scriptstyle \blacksquare}$}
 \put(221,118){${\scriptstyle \blacksquare}$}
 \put(206,123){${\scriptstyle \blacksquare}$}
 \put(211,123){${\scriptstyle \blacksquare}$}
 \put(216,123){${\scriptstyle \blacksquare}$}
 \put(221,123){${\scriptstyle \blacksquare}$}
 \put(206,128){${\scriptstyle \blacksquare}$}
 \put(211,128){${\scriptstyle \blacksquare}$}
 \put(216,128){${\scriptstyle \blacksquare}$}
 \put(221,128){${\scriptstyle \blacksquare}$}
 \put(206,133){${\scriptstyle \blacksquare}$}
 \put(211,133){${\scriptstyle \blacksquare}$}
 \put(216,133){${\scriptstyle \blacksquare}$}
 \put(221,133){${\scriptstyle \blacksquare}$}
 \put(206,138){${\scriptstyle \blacksquare}$}
 \put(211,138){${\scriptstyle \blacksquare}$}
 \put(216,138){${\scriptstyle \blacksquare}$}
 \put(221,138){${\scriptstyle \blacksquare}$}
 \put(226,123){${\scriptstyle \blacksquare}$}
 \put(231,133){${\textstyle \textbf{p}}$}
 \qbezier(216,113)(217,111)(222,101)
 \put(222,101){\circle*{2}}
 \qbezier[20](222,101)(228,93)(231,83)
 \put(231,83){\circle*{2}}
 \qbezier(231,83)(232,81)(237,71)
 \put(237,71){\circle*{2}}
 \put(232,42){${\scriptstyle \blacksquare}$}
 \put(237,42){${\scriptstyle \blacksquare}$}
 \put(242,42){${\scriptstyle \blacksquare}$}
 \put(247,42){${\scriptstyle \blacksquare}$}
 \put(252,42){${\scriptstyle \blacksquare}$}
 \put(232,47){${\scriptstyle \blacksquare}$}
 \put(237,47){${\scriptstyle \blacksquare}$}
 \put(242,47){${\scriptstyle \blacksquare}$}
 \put(247,47){${\scriptstyle \blacksquare}$}
 \put(252,47){${\scriptstyle \blacksquare}$}
 \put(232,52){${\scriptstyle \blacksquare}$}
 \put(237,52){${\scriptstyle \blacksquare}$}
 \put(242,52){${\scriptstyle \blacksquare}$}
 \put(247,52){${\scriptstyle \blacksquare}$}
 \put(252,52){${\scriptstyle \blacksquare}$}
 \put(232,57){${\scriptstyle \blacksquare}$}
 \put(237,57){${\scriptstyle \blacksquare}$}
 \put(242,57){${\scriptstyle \blacksquare}$}
 \put(247,57){${\scriptstyle \blacksquare}$}
 \put(252,57){${\scriptstyle \blacksquare}$}
 \put(232,62){${\scriptstyle \blacksquare}$}
 \put(237,62){${\scriptstyle \blacksquare}$}
 \put(242,62){${\scriptstyle \blacksquare}$}
 \put(247,62){${\scriptstyle \blacksquare}$}
 \put(252,62){${\scriptstyle \blacksquare}$}
 \put(242,34){${\scriptstyle \ell_{\textrm{c}}}$}
 \put(260,47){${\scriptstyle \ell_{\textrm{c}}}$}
 \put(237,71){\line(1,2){20}}
 \put(257,111){\circle*{2}}
 \put(250,116){${\scriptstyle \blacksquare}$}
 \put(255,116){${\scriptstyle \blacksquare}$}
 \put(260,116){${\scriptstyle \blacksquare}$}
 \put(265,116){${\scriptstyle \blacksquare}$}
 \put(270,116){${\scriptstyle \blacksquare}$}
 \put(250,121){${\scriptstyle \blacksquare}$}
 \put(255,121){${\scriptstyle \blacksquare}$}
 \put(260,121){${\scriptstyle \blacksquare}$}
 \put(265,121){${\scriptstyle \blacksquare}$}
 \put(270,121){${\scriptstyle \blacksquare}$}
 \put(250,126){${\scriptstyle \blacksquare}$}
 \put(255,126){${\scriptstyle \blacksquare}$}
 \put(260,126){${\scriptstyle \blacksquare}$}
 \put(265,126){${\scriptstyle \blacksquare}$}
 \put(270,126){${\scriptstyle \blacksquare}$}
 \put(250,131){${\scriptstyle \blacksquare}$}
 \put(255,131){${\scriptstyle \blacksquare}$}
 \put(260,131){${\scriptstyle \blacksquare}$}
 \put(265,131){${\scriptstyle \blacksquare}$}
 \put(270,131){${\scriptstyle \blacksquare}$}
 \put(250,136){${\scriptstyle \blacksquare}$}
 \put(255,136){${\scriptstyle \blacksquare}$}
 \put(260,136){${\scriptstyle \blacksquare}$}
 \put(265,136){${\scriptstyle \blacksquare}$}
 \put(270,136){${\scriptstyle \blacksquare}$}
 \put(275,121){${\scriptstyle \blacksquare}$}
 \qbezier[20](257,111)(260,105)(263,99)
 \put(155,90){${\scriptstyle 4-2h}$}
 \put(185,100){${\scriptstyle 2h}$}
 \end{picture}
\caption{Schematical representation of the optimal path between 
$\textbf{d}$ and $\textbf{u}$ for the 
Ising model with indication of energy differences. 
Black squares represent pluses.}
\label{f:fig04}
\end{figure}

We denote by $\textbf{d}$ and $\textbf{u}$ the two homogeneous 
configurations with spins respectively equal to $-1$ and $+1$. 
It is not surprising that 
$\Omega^\textrm{s}=\{\textbf{u}\}$ and 
$\Omega^\textrm{m}=\{\textbf{d}\}$, provided the Hamiltonian 
is the standard 2D Ising Hamiltonian defined in 
\eqref{isi000}.
The optimal path connecting $\textbf{d}$ to $\textbf{u}$ 
is depicted in figure~\ref{f:fig04}:
four minuses are flipped to plus one after the other to form a two
by two square, the sides of the droplet grow one after the other in
order to obtain alternately a rectangle of pluses with side length
difference equal to one and a square of pluses till $\textbf{u}$
is reached.
A side is added to a rectangle (or a square) by flipping 
to plus a minus spin adjacent to the rectangle (adding one 
protuberance) and then flipping, one after the other, 
minus spins with two adjacent pluses until a rectangular or 
square shape is recovered. 

When a plus protuberance is added to the side of the plus rectangle (or
square) of length $\ell$ the energy increases by $4-2h$; on the other hand, 
when this very side is filled by pluses 
the energy decreases by $2h(\ell-1)$. 
The energy difference related to the process of adding a side is 
equal to $4-2h-2h(\ell-1)$, which is positive provided $\ell<\ell_\textup{c}$, 
where 
$\ell_\textup{c}=\lfloor2/h\rfloor+1$,
where, for any real number $a$, $\lfloor a\rfloor$ is its 
integer part, i.e., the largest integer smaller than $a$. 
Thus the optimal path achieves its maximal energy 
at the configuration $\textbf{p}$, see figure~\ref{f:fig04},
made of a $\ell_\textup{c}(\ell_\textup{c}-1)$ 
rectangle of pluses with a unit protuberance 
on one of its longest sides and 
the communication height between $\textbf{d}$ and $\textbf{u}$
is
\begin{equation}
\label{misi010}
\Gamma
=
H(\textbf{p})-H(\textbf{d})
=
8\ell_\textup{c}-2h[\ell_\textup{c}(\ell_\textup{c}-1)+1]
\sim
\frac{8}{h}
,
\end{equation}
where $H$ is the Ising Hamiltonian \eqref{isi000} and 
the estimate is valid for $h\to0$. 

With the model dependent ingredients summarized in this section, 
the results reported in Section~\ref{s:generale} provide a 
full description of the metastable behavior of the Ising 
model: the state $\textbf{d}$ is metastable and, if the system 
is prepared in $\textbf{d}$, at small temperature 
the typical time necessary to 
reach the stable state $\textbf{u}$ is of order 
$\exp\{8/(hT)\}$. 
Moreover, during the transition from the metastable to the stable 
state, with high probability, the system visits the configuration 
$\textbf{p}$. 
This last remark is very important, since it means that 
the transition to the stable state is performed through 
the sudden nucleation of a large critical droplet and not 
via the coalescence of many small droplets distributed 
throughout the whole volume $\Lambda$. 

\subsection{Metastable behavior of the reversible nearest--neighbor PCA}
\label{s:met-pca} 
\par\noindent
We consider the synchronous dynamics (PCA) introduced 
in Section~\ref{s:sincrono} for the standard 2D Ising Hamiltonian 
considered in Section~\ref{s:ising} with $0<h\ll 1$ such that 
$2/h$ is not integer. We also assume that the side length 
of the lattice $\Lambda$ is an even number.
The metastable behavior for 
such a model was first studied in \cite{cirillo2003metastability};
we describe 
here the main results using the language developed above.

As in the previous section, 
we denote by $\textbf{d}$ and $\textbf{u}$ the two homogeneous 
configurations with spins respectively equal to $-1$ and $+1$. 
Moreover, we consider the two chessboard configurations 
such that all the spins associated with sites on the 
even sub--lattice are equal and opposite to the spins 
associated with sites on the odd sub--lattice. These two 
identified configurations are denoted by
$\textbf{c}$. The two chessboards are identified, since, with high 
probability at low temperature, the chain started at one of them 
is trapped in a continuous flip--flop among the two which 
is performed with no energy cost. 
The presence of this flip--flopping configurations is a 
signature of the parallel (synchronous)  character of the 
dynamics.
It is proven in \cite{cirillo2003metastability} that
$\Omega^\textrm{s}=\{\textbf{u}\}$ and 
$\Omega^\textrm{m}=\{\textbf{d},\textbf{c}\}$, with respect 
to the Hamiltonian \eqref{isi005}.
The fact that chessboard configurations play a crucial
role in the metastable behavior of the model is also due 
to the antiferromagnetic term $J_\diamond$ which is present 
in the Hamiltonian \eqref{isi010}.

\setlength{\unitlength}{1.2pt}
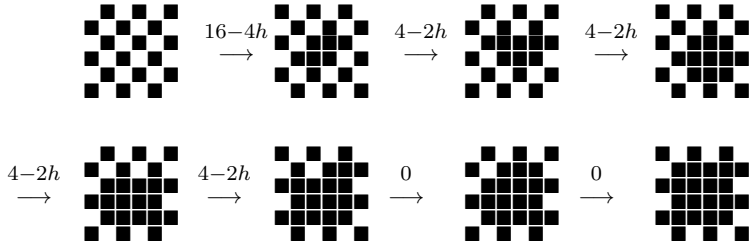
\begin{figure}
 \begin{picture}(300,100)(-90,-60)
 \thinlines
 \put(0,0){${\scriptstyle \blacksquare}$}
 \put(10,0){${\scriptstyle \blacksquare}$}
 \put(20,0){${\scriptstyle \blacksquare}$}
 \put(5,5){${\scriptstyle \blacksquare}$}
 \put(15,5){${\scriptstyle \blacksquare}$}
 \put(25,5){${\scriptstyle \blacksquare}$}
 \put(0,10){${\scriptstyle \blacksquare}$}
 \put(10,10){${\scriptstyle \blacksquare}$}
 \put(20,10){${\scriptstyle \blacksquare}$}
 \put(5,15){${\scriptstyle \blacksquare}$}
 \put(15,15){${\scriptstyle \blacksquare}$}
 \put(25,15){${\scriptstyle \blacksquare}$}
 \put(0,20){${\scriptstyle \blacksquare}$}
 \put(10,20){${\scriptstyle \blacksquare}$}
 \put(20,20){${\scriptstyle \blacksquare}$}
 \put(5,25){${\scriptstyle \blacksquare}$}
 \put(15,25){${\scriptstyle \blacksquare}$}
 \put(25,25){${\scriptstyle \blacksquare}$}
\put(37,15){$\newatop{\scriptstyle 16-4h}{\longrightarrow}$}
 \put(65,0){${\scriptstyle \blacksquare}$}
 \put(75,0){${\scriptstyle \blacksquare}$}
 \put(85,0){${\scriptstyle \blacksquare}$}
 \put(60,5){${\scriptstyle \blacksquare}$}
 \put(70,5){${\scriptstyle \blacksquare}$}
 \put(80,5){${\scriptstyle \blacksquare}$}
 \put(65,10){${\scriptstyle \blacksquare}$}
 \put(70,10){${\scriptstyle \blacksquare}$}
 \put(75,10){${\scriptstyle \blacksquare}$}
 \put(85,10){${\scriptstyle \blacksquare}$}
 \put(60,15){${\scriptstyle \blacksquare}$}
 \put(70,15){${\scriptstyle \blacksquare}$}
 \put(75,15){${\scriptstyle \blacksquare}$}
 \put(80,15){${\scriptstyle \blacksquare}$}
 \put(65,20){${\scriptstyle \blacksquare}$}
 \put(75,20){${\scriptstyle \blacksquare}$}
 \put(85,20){${\scriptstyle \blacksquare}$}
 \put(60,25){${\scriptstyle \blacksquare}$}
 \put(70,25){${\scriptstyle \blacksquare}$}
 \put(80,25){${\scriptstyle \blacksquare}$}
\put(97,15){$\newatop{\scriptstyle 4-2h}{\longrightarrow}$}
 \put(120,0){${\scriptstyle \blacksquare}$}
 \put(130,0){${\scriptstyle \blacksquare}$}
 \put(140,0){${\scriptstyle \blacksquare}$}
 \put(125,5){${\scriptstyle \blacksquare}$}
 \put(135,5){${\scriptstyle \blacksquare}$}
 \put(145,5){${\scriptstyle \blacksquare}$}
 \put(120,10){${\scriptstyle \blacksquare}$}
 \put(130,10){${\scriptstyle \blacksquare}$}
 \put(135,10){${\scriptstyle \blacksquare}$}
 \put(140,10){${\scriptstyle \blacksquare}$}
 \put(125,15){${\scriptstyle \blacksquare}$}
 \put(130,15){${\scriptstyle \blacksquare}$}
 \put(135,15){${\scriptstyle \blacksquare}$}
 \put(140,15){${\scriptstyle \blacksquare}$}
 \put(145,15){${\scriptstyle \blacksquare}$}
 \put(120,20){${\scriptstyle \blacksquare}$}
 \put(130,20){${\scriptstyle \blacksquare}$}
 \put(140,20){${\scriptstyle \blacksquare}$}
 \put(125,25){${\scriptstyle \blacksquare}$}
 \put(135,25){${\scriptstyle \blacksquare}$}
 \put(145,25){${\scriptstyle \blacksquare}$}
\put(157,15){$\newatop{\scriptstyle 4-2h}{\longrightarrow}$}
 \put(185,0){${\scriptstyle \blacksquare}$}
 \put(195,0){${\scriptstyle \blacksquare}$}
 \put(205,0){${\scriptstyle \blacksquare}$}
 \put(180,5){${\scriptstyle \blacksquare}$}
 \put(190,5){${\scriptstyle \blacksquare}$}
 \put(195,5){${\scriptstyle \blacksquare}$}
 \put(200,5){${\scriptstyle \blacksquare}$}
 \put(185,10){${\scriptstyle \blacksquare}$}
 \put(190,10){${\scriptstyle \blacksquare}$}
 \put(195,10){${\scriptstyle \blacksquare}$}
 \put(200,10){${\scriptstyle \blacksquare}$}
 \put(205,10){${\scriptstyle \blacksquare}$}
 \put(180,15){${\scriptstyle \blacksquare}$}
 \put(190,15){${\scriptstyle \blacksquare}$}
 \put(195,15){${\scriptstyle \blacksquare}$}
 \put(200,15){${\scriptstyle \blacksquare}$}
 \put(185,20){${\scriptstyle \blacksquare}$}
 \put(195,20){${\scriptstyle \blacksquare}$}
 \put(205,20){${\scriptstyle \blacksquare}$}
 \put(180,25){${\scriptstyle \blacksquare}$}
 \put(190,25){${\scriptstyle \blacksquare}$}
 \put(200,25){${\scriptstyle \blacksquare}$}
\put(-25,-30){$\newatop{\scriptstyle 4-2h}{\longrightarrow}$}
 \put(5,-20){${\scriptstyle \blacksquare}$}
 \put(15,-20){${\scriptstyle \blacksquare}$}
 \put(25,-20){${\scriptstyle \blacksquare}$}
 \put(0,-25){${\scriptstyle \blacksquare}$}
 \put(10,-25){${\scriptstyle \blacksquare}$}
 \put(20,-25){${\scriptstyle \blacksquare}$}
 \put(5,-30){${\scriptstyle \blacksquare}$}
 \put(10,-30){${\scriptstyle \blacksquare}$}
 \put(15,-30){${\scriptstyle \blacksquare}$}
 \put(20,-30){${\scriptstyle \blacksquare}$}
 \put(25,-30){${\scriptstyle \blacksquare}$}
 \put(0,-35){${\scriptstyle \blacksquare}$}
 \put(5,-35){${\scriptstyle \blacksquare}$}
 \put(10,-35){${\scriptstyle \blacksquare}$}
 \put(15,-35){${\scriptstyle \blacksquare}$}
 \put(20,-35){${\scriptstyle \blacksquare}$}
 \put(5,-40){${\scriptstyle \blacksquare}$}
 \put(10,-40){${\scriptstyle \blacksquare}$}
 \put(15,-40){${\scriptstyle \blacksquare}$}
 \put(20,-40){${\scriptstyle \blacksquare}$}
 \put(25,-40){${\scriptstyle \blacksquare}$}
 \put(0,-45){${\scriptstyle \blacksquare}$}
 \put(10,-45){${\scriptstyle \blacksquare}$}
 \put(20,-45){${\scriptstyle \blacksquare}$}
\put(35,-30){$\newatop{\scriptstyle 4-2h}{\longrightarrow}$}
 \put(60,-20){${\scriptstyle \blacksquare}$}
 \put(70,-20){${\scriptstyle \blacksquare}$}
 \put(80,-20){${\scriptstyle \blacksquare}$}
 \put(65,-25){${\scriptstyle \blacksquare}$}
 \put(75,-25){${\scriptstyle \blacksquare}$}
 \put(80,-25){${\scriptstyle \blacksquare}$}
 \put(85,-25){${\scriptstyle \blacksquare}$}
 \put(60,-30){${\scriptstyle \blacksquare}$}
 \put(65,-30){${\scriptstyle \blacksquare}$}
 \put(70,-30){${\scriptstyle \blacksquare}$}
 \put(75,-30){${\scriptstyle \blacksquare}$}
 \put(80,-30){${\scriptstyle \blacksquare}$}
 \put(65,-35){${\scriptstyle \blacksquare}$}
 \put(70,-35){${\scriptstyle \blacksquare}$}
 \put(75,-35){${\scriptstyle \blacksquare}$}
 \put(80,-35){${\scriptstyle \blacksquare}$}
 \put(85,-35){${\scriptstyle \blacksquare}$}
 \put(60,-40){${\scriptstyle \blacksquare}$}
 \put(65,-40){${\scriptstyle \blacksquare}$}
 \put(70,-40){${\scriptstyle \blacksquare}$}
 \put(75,-40){${\scriptstyle \blacksquare}$}
 \put(80,-40){${\scriptstyle \blacksquare}$}
 \put(65,-45){${\scriptstyle \blacksquare}$}
 \put(75,-45){${\scriptstyle \blacksquare}$}
 \put(85,-45){${\scriptstyle \blacksquare}$}
\put(95,-30){$\newatop{\scriptstyle 0}{\longrightarrow}$}
 \put(125,-20){${\scriptstyle \blacksquare}$}
 \put(135,-20){${\scriptstyle \blacksquare}$}
 \put(145,-20){${\scriptstyle \blacksquare}$}
 \put(120,-25){${\scriptstyle \blacksquare}$}
 \put(130,-25){${\scriptstyle \blacksquare}$}
 \put(135,-25){${\scriptstyle \blacksquare}$}
 \put(140,-25){${\scriptstyle \blacksquare}$}
 \put(125,-30){${\scriptstyle \blacksquare}$}
 \put(130,-30){${\scriptstyle \blacksquare}$}
 \put(135,-30){${\scriptstyle \blacksquare}$}
 \put(140,-30){${\scriptstyle \blacksquare}$}
 \put(145,-30){${\scriptstyle \blacksquare}$}
 \put(120,-35){${\scriptstyle \blacksquare}$}
 \put(125,-35){${\scriptstyle \blacksquare}$}
 \put(130,-35){${\scriptstyle \blacksquare}$}
 \put(135,-35){${\scriptstyle \blacksquare}$}
 \put(140,-35){${\scriptstyle \blacksquare}$}
 \put(125,-40){${\scriptstyle \blacksquare}$}
 \put(130,-40){${\scriptstyle \blacksquare}$}
 \put(135,-40){${\scriptstyle \blacksquare}$}
 \put(140,-40){${\scriptstyle \blacksquare}$}
 \put(145,-40){${\scriptstyle \blacksquare}$}
 \put(120,-45){${\scriptstyle \blacksquare}$}
 \put(130,-45){${\scriptstyle \blacksquare}$}
 \put(140,-45){${\scriptstyle \blacksquare}$}
\put(155,-30){$\newatop{\scriptstyle 0}{\longrightarrow}$}
 \put(180,-20){${\scriptstyle \blacksquare}$}
 \put(190,-20){${\scriptstyle \blacksquare}$}
 \put(200,-20){${\scriptstyle \blacksquare}$}
 \put(185,-25){${\scriptstyle \blacksquare}$}
 \put(190,-25){${\scriptstyle \blacksquare}$}
 \put(195,-25){${\scriptstyle \blacksquare}$}
 \put(200,-25){${\scriptstyle \blacksquare}$}
 \put(205,-25){${\scriptstyle \blacksquare}$}
 \put(180,-30){${\scriptstyle \blacksquare}$}
 \put(185,-30){${\scriptstyle \blacksquare}$}
 \put(190,-30){${\scriptstyle \blacksquare}$}
 \put(195,-30){${\scriptstyle \blacksquare}$}
 \put(200,-30){${\scriptstyle \blacksquare}$}
 \put(185,-35){${\scriptstyle \blacksquare}$}
 \put(190,-35){${\scriptstyle \blacksquare}$}
 \put(195,-35){${\scriptstyle \blacksquare}$}
 \put(200,-35){${\scriptstyle \blacksquare}$}
 \put(205,-35){${\scriptstyle \blacksquare}$}
 \put(180,-40){${\scriptstyle \blacksquare}$}
 \put(185,-40){${\scriptstyle \blacksquare}$}
 \put(190,-40){${\scriptstyle \blacksquare}$}
 \put(195,-40){${\scriptstyle \blacksquare}$}
 \put(200,-40){${\scriptstyle \blacksquare}$}
 \put(185,-45){${\scriptstyle \blacksquare}$}
 \put(195,-45){${\scriptstyle \blacksquare}$}
 \put(205,-45){${\scriptstyle \blacksquare}$}
\end{picture}
\caption{Schematical representation of the optimal path between 
$\textbf{c}$ and $\textbf{u}$ for the 
PCA model with indication of the energy cost computed using 
\eqref{asi060} with $I$ the set of nearest neighbors and $J_{ij}=1$ 
for $i$ and $j$ nearest neighbors.
Black squares represent pluses, white ones minuses.}
\label{f:fig05}
\end{figure}

The optimal path connecting $\textbf{c}$ to $\textbf{u}$ 
is constructed, see also 
\cite[Section~5.2, case A$_3$--A$_4$]{bet2021effect},
by letting the spins to perform a flip--flop 
at each time except for some pluses that are kept fixed 
in such a way to eventually 
invade the whole lattice (see figure~\ref{f:fig05}): 
at the first step two arbitrary next--to--the--nearest 
pluses (Euclidean distance $\sqrt{2}$) are kept fixed, so that 
a two by two plus square is formed in the chessboard sea. 
At the second step the pluses in the square and 
a plus adjacent to the square is kept fixed 
in order to form a two by three rectangle.
At the third step the pluses in the rectangle are kept fixed 
together with the plus at the center of one chessboard side 
adjacent to the longest side of the rectangle. In this way a 
three by three plus rectangle is formed. 
Before taking the fourth step the chessboard sea is let flip--flop,
if necessary, so 
that a plus appears at the site in the center of
one of the chessboard sides adjacent to the plus square, thus such a
spin is kept fixed and a three times four rectangle is formed.
In the fifth step one of the pluses adjacent to the 
longest side of the plus rectangle is kept fixed and a double 
or triple plus protuberance is formed on such a side. In the following
steps this protuberance is kept fixed together with the plus in the 
three by four rectangle till the four by four plus rectangle is formed.
This path is then followed alike alternating plus squares to plus 
rectangles whose side lengths differ by one.

When one plus adjacent from the exterior to a square or a rectangle of 
pluses is kept during the flip--flop, the energy cost is equal to
$4-2h$. This is the total cost payed to add a plus slice, since the 
following steps have no cost. On the way back, eroding a plus slice 
of length $\ell$ has a cost $2h(\ell-1)$ 
since at each step the external pluses,
having only two neighboring pluses, are flipped paying the cost 2h
The last step is cost free, because the last
plus has just one neighboring plus. 
By repeating the computation performed in the asynchronous case, 
one finds again the critical length 
$\ell_\textup{c}=\lfloor2/h\rfloor+1$ for the plus droplets
in the chessboard sea. 
Thus the optimal path achieves its maximal height in the 
jump from the $\ell_\textup{c}\times(\ell_\textup{c}-1)$ 
plus rectangle
to the configuration in which a double or triple plus protuberance 
is added to the slab adjacent to the longest side of the same 
rectangle. Finally the communication height between $\textbf{c}$ and 
$\textbf{u}$ is 
\begin{equation}
\label{mpca000}
\Gamma
=
H(\textbf{q}_2)-H(\textbf{c})+2h(\ell_\textup{c}-1)
=
-2h\ell_\textup{c}^2+(8+2h)\ell_\textup{c}+4h
\sim
\frac{8}{h}
,
\end{equation}
where
$\textbf{q}_2$ is the $\ell_\textup{c}\times\ell_\textup{c}$ plus droplet in 
the sea of chessboard, 
$H$ is the Hamiltonian \eqref{isi005} and 
the estimate is valid for $h\to0$. 

We do not describe in detail the optimal path between $\textbf{d}$ and 
$\textbf{u}$, we just mention that it is made of two 
parts: the first part from $\textbf{d}$ to $\textbf{c}$ realizes
the growth of the chessboard square droplet $\textbf{q}_1$
of size $\lfloor2/h\rfloor+1$ 
in the sea of minuses and its eventual growth to $\textbf{c}$,
see \cite[Section~5.2, case A$_1$--A$_2$]{bet2021effect}, the second 
part is precisely the optimal path from $\textbf{c}$ to $\textbf{u}$, 
see also \cite[Section~5.2, case A$_3$--A$_4$]{bet2021effect}.
Its maximal height is equal to the value $\Gamma$ computed in 
\eqref{mpca000} and this explains why this model have two 
metastable states. 

With the model dependent ingredients summarized in this section, 
the results reported in Section~\ref{s:generale} provide a 
full description of the metastable behavior of the Ising 
model: the states $\textbf{d}$ and $\textbf{c}$ 
are metastable and, if the system 
is prepared in $\textbf{d}$, at small temperature 
the typical time necessary to 
reach the stable state $\textbf{u}$ is of order 
$\exp\{8/(hT)\}$. 
Moreover, during the transition from the metastable to the stable 
state, with high probability, the system visits the configurations
$\textbf{q}_1$, $\textbf{c}$, and $\textbf{q}_2$.

This situation has been called in the literature as ``series of 
metastable states". We refer the interested reader
to the papers 
\cite{cirillo2017sum} for some general results and their 
applications to the case of the Blume--Capel model with zero chemical
potential, \cite{cirillo2016sum} for the discussion of this phenomenon 
in the context of Probabilistic Cellular Automata, and to 
\cite{bet2021effect} for the extension of these results to an arbitrary
series of metastable states. 
The case of the Blume--Capel model is also discussed in 
\cite{landimlemire}.

\subsection{Numerical simulations}
\label{s:simul} 
\par\noindent
We illustrate by means of numerical simulations the nucleation phenomenon
described on a rigorous basis above. 
We refer to \cite{RikvoldTomitaetal} for a detailed numerical study of 
the exit time for the 2D Ising model. 

\begin{figure}
\begin{center}
\includegraphics[width=0.12\textwidth]{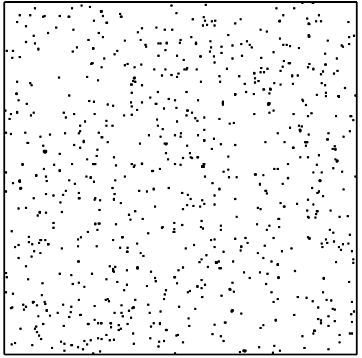}
\hskip 0.2 cm
\includegraphics[width=0.12\textwidth]{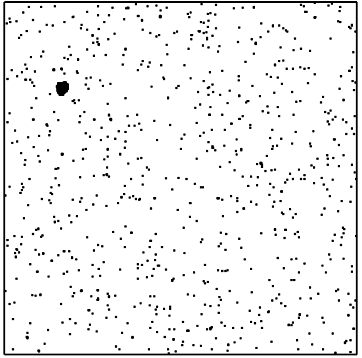}
\hskip 0.2 cm
\includegraphics[width=0.12\textwidth]{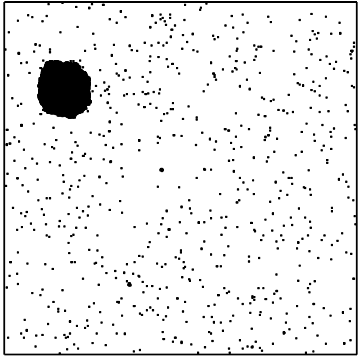}
\hskip 0.2 cm
\includegraphics[width=0.12\textwidth]{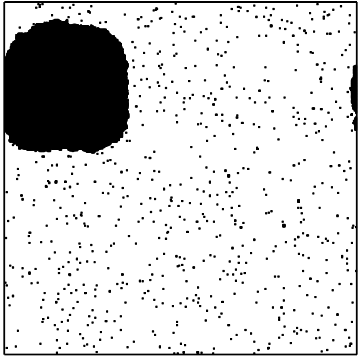}
\hskip 0.2 cm
\includegraphics[width=0.12\textwidth]{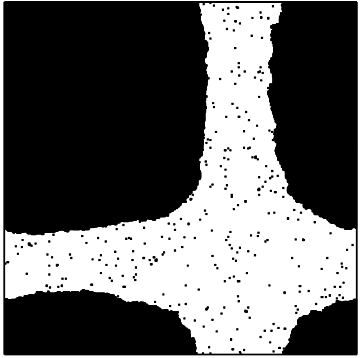}
\end{center}
\caption{Nucleation of the plus phase in the asynchronous Ising 
model with $h=0.2$ and $1/T=0.78$. 
From the left to the right the configurations after 
$800$, $1000$, $1400$, $2000$, and $3200$ Monte Carlo
full lattice sweeps of the lattice are reported.
Black and white spots correspond to plus and minus spins, respectively.}
\label{f:fig06}
\end{figure}

We first consider the asynchronous model and we simulate the 
stochastic Ising model on the $512\times512$ lattice 
with magnetic field $h=0.2$. In figure~\ref{f:fig06} we show 
the nucleation of the plus phase starting from the minus 
metastable state at inverse temperature $1/T=0.78$. 
The sequence of configurations shows that the nucleation is 
performed via the formation of a single droplet of pluses in the 
sea of minuses. Due to the fact that in the simulation we could 
not consider a too small value of $T$, the droplet is not perfectly 
rectangular and the sea of minuses is full of very small 
sub--critical droplets of pluses. 

\begin{figure}
\begin{center}
\includegraphics[width=0.12\textwidth]{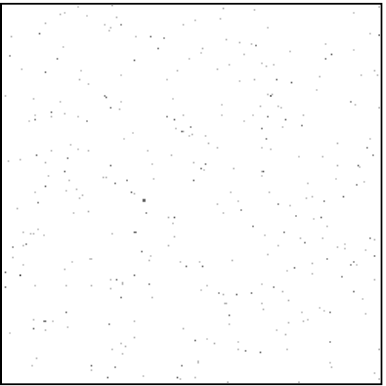}
\hskip 0.2 cm
\includegraphics[width=0.12\textwidth]{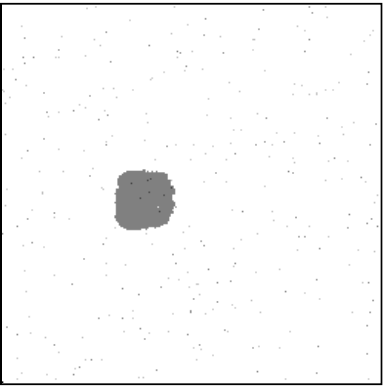}
\hskip 0.2 cm
\includegraphics[width=0.12\textwidth]{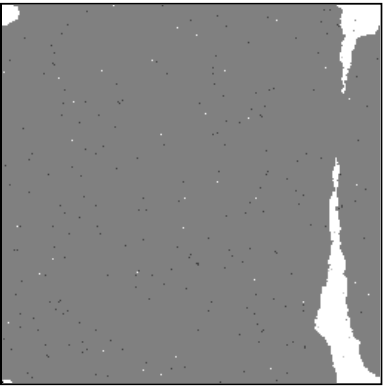}
\hskip 0.2 cm
\includegraphics[width=0.12\textwidth]{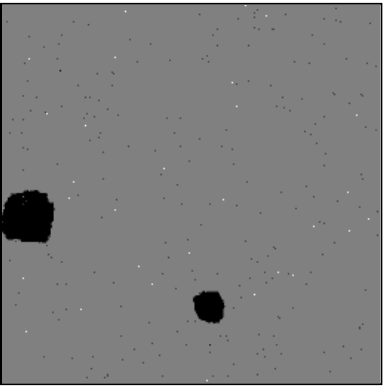}
\hskip 0.2 cm
\includegraphics[width=0.12\textwidth]{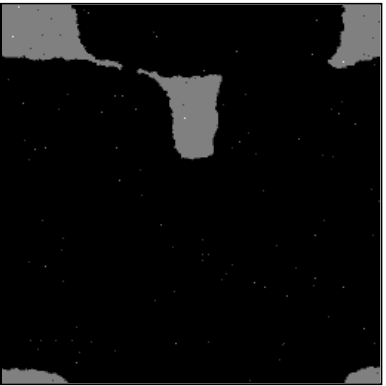}
\end{center}
\caption{Nucleation of the plus phase in the synchronous nearest
neighbor PCA 
model with $h=0.3$ and $1/T=0.90$. 
From the left to the right the configurations after 
$3000$, $3600$, $6000$, $7600$, and $9600$ Monte Carlo
full lattice sweeps of the lattice are reported.
Configurations on $2\times2$ tiles are reported using grayscale 
from $-4$ (white) to $+4$ (black).}
\label{f:fig07}
\end{figure}

In figure~\ref{f:fig07} we illustrate the more complex
nucleation phenomenon for the asynchronous nearest neighbor 
PCA model on the $512\times512$ lattice 
with magnetic field $h=0.3$ and $1/T=0.90$. In order to show the nucleation 
of the chessboard configuration, we represent the whole configuration 
on the $256\times256$ lattice and we associate with each site of the 
new lattice the sum of the spins in a $2\times2$ tile of the original 
lattice. The new block variables, taking 
values in $\{-4,-2, 0,+2,+4\}$, are plotted using a grayscale 
paddle ranging from white ($-4$) to black ($+4$). 
The sequence of configurations shows that the nucleation of the 
chessboard (gray) configuration is
performed via the formation of a single droplet in the 
sea of minuses. Then, the plus phase is nucleated in the sea 
of chessboard via the formation of a droplet of pluses. Due to the 
relative high value of the temperature that we used in the simulations 
the formation of two plus droplets is observed.  

\begin{figure}
\begin{center}
\includegraphics[width=0.235\textwidth]{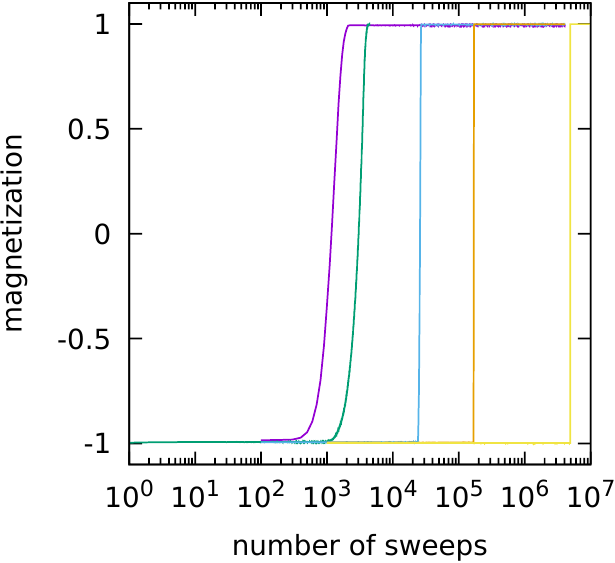}
\hskip .08 cm
\includegraphics[width=0.235\textwidth]{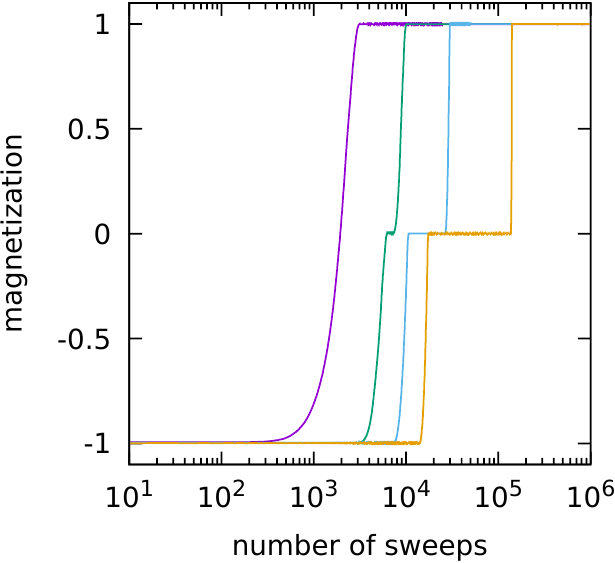}
\hskip .08 cm
\includegraphics[width=0.239\textwidth]{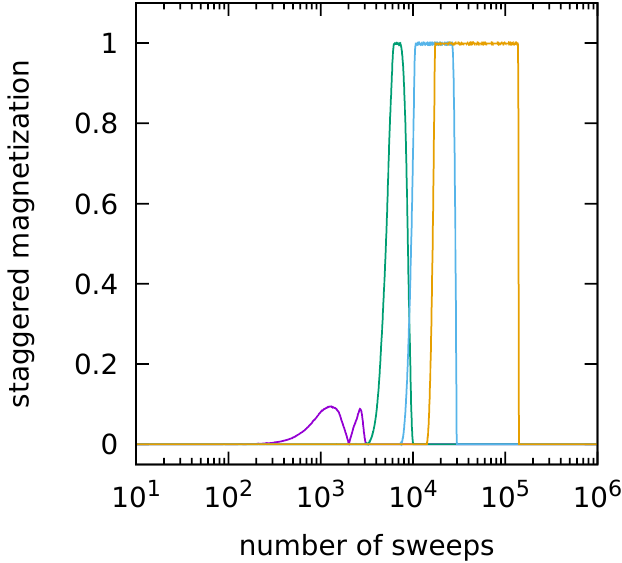}
\end{center}
\caption{Magnetization and staggered magnetization 
versus number of full Monte Carlo lattice 
sweeps. 
Left: magnetization of the asynchronous Ising model on the 
$512\times512$ lattice 
with $h=0.2$ and $1/T=0.70,0.79,0.80,0.85,0.90$ (respectively from the 
left to the right, i.e., purple, green, blue, orange, yellow).
Center: magnetizations of the 
synchronous nearest neighbor reversible PCA on the $512\times512$ lattice 
with $h=0.3$ and $1/T=0.85,0.90,0.95,1.00$ (respectively from the 
left to the right, i.e., purple, green, blue, orange).
Right: staggered magnetization for the same model as in the 
center panel. 
}
\label{f:fig08}
\end{figure}

In order to give a vivid idea of the time scales involved in the 
nucleation process, we plot in figure~\ref{f:fig08} the magnetization,
namely, the sum of all the spins on the lattice, versus the number 
of full Monte Carlo sweeps for the synchronous and asynchronous dynamics 
started at the minus configuration. 
A full Monte Carlo sweep corresponds 
to one step of the dynamics in the synchronous case and to a sequence 
of steps equal to the number of sites of the lattice in the 
asynchronous case. 

It is not reasonable to compare the time scales observed 
in the figure to the result \eqref{gen070} which is an exact 
prediction for the exit time in the limit $T\to0$. Indeed, 
in simulating metastability effects it is necessary to use 
not very small values of the temperature, otherwise the 
dynamics would stick to the metastable state and nothing would be observed.
We refer to \cite[Figure~3]{cirillo1998metastability} for the numerical
estimate of the exit time for the Ising model with free boundary 
conditions on a small lattice. 

Nevertheless, it is very interesting to remark how neatly the simulations
show, in both cases, that the exit time from the metastable state 
increases dramatically when $T$ becomes smaller and smaller (note that 
in the figure the horizontal scale is logarithmic). 
Moreover, the data in the pictures also show that the exit from the 
metastable state is an abrupt phenomenon: the system is trapped in the 
metastable state for a long time, but when it performs the transition, 
this process is completed in a very small time.  

It is also very interesting 
to observe that, in the PCA case, the dynamics 
started at the minus configuration, before reaching the plus state, 
spends a huge time in a configuration with zero magnetization, 
namely, the chessboard metastable state. This effect is lost if the 
temperature is so large (purple curve) that fluctuations dominate 
with respect to the behavior driven by the energy landscape.
The fact that the intermediate zero magnetization 
state is precisely the flip--flopping chessboard
configuration is demonstrated by the right panel where we plotted the 
staggered magnetization, i.e., the absolute value of the difference 
between the sum of the spins on the even sub--lattice and the sum
of the spin on the odd one, versus the number of full sweeps

\section{Concluding remarks}
\label{s:concusioni} 
\par\noindent
We have shown that starting from the same spin system Hamiltonian 
\eqref{mod000} depending on how the spin changes are 
implemented, asynchronously or synchronously, two different 
stochastic models are found. The two models differ not only
for their stationary behavior, i.e., the stationary Gibbs measure, 
but also for their dynamical behavior. 
In particular we have discussed these dynamical aspects in the framework 
of the metastability theory and,
in this connection, we have shown that, depending on how
the dynamics is implemented, different metastable states 
can be found. Moreover, the critical configurations and the 
exit time also turn out to be different. 

In this review we have compared the asynchronous heat bath dynamics for the Ising model with its natural parallel implementation.  However, one could have considered the parallel implementation of an asynchronous Metropolis dynamics, as done in \cite{cirillo2002}. Here indeed a  2D Ising model 
with Hamiltonian \eqref{isi000} is considered and  the spins
on the even  (odd)  sub--lattice  are  simultaneously  updated by using the Metropolis weights  \cite{hollander2000metastability}. Given a random set of sites $I$ belonging to the even (odd) sub--lattice, 
and sampled under $\nu$,  the transition matrix is given by
\begin{equation}
\label{par02}
p_T(\sigma,\eta)=
\left\{
\begin{array}{ll}
\!\!
\frac{\nu(I)}{|\Lambda|} e^{- [H(\eta)-H(\sigma)]_+/T}
&
\!\!
\textup{if }
\eta(j)=-\sigma(j)  \textup{ for } j\in I, 
\eta(j)=\sigma(j) \textup{ for } j\in \Lambda\setminus I, 
\\
\!\!
0
&
\!\!
\textup{otherwise}
\end{array}
\right.
,
\end{equation}
where, 
for any $a\in\mathbb{R}$, the positive part $[a]_+$ of $a$ 
is equal to $a$ if $a>0$ and to $0$ otherwise.
The Markov chain defined by (\ref{par02}) is reversible with respect 
to the Gibbs measure defined in (\ref{sin020}).
Despite the fact that  new jumps have been allowed  by the dynamics, in \cite{cirillo2002} has been shown that the metastable character of the one--site Metropolis dynamics is preserved. The key observation is that 
 the flip of the spins in $I$ can be substituted by a suitable sequence of  sequential single spin flip events. Since the sites in $I$
 belong to the same sub--lattice, the author could prove that a configuration
 is a local minimum for the multi--spin dynamics if and only if it is a
local minimum in the single spin flip case. A natural generalization of the model might be to consider a general set $I$, not necessarily belonging to a sub--lattice. In this case the set $I$ might be decomposed in a connected 
 (in the sense of nearest neighbors) part and in a non--connected part. The non--connected part might behave as the case in \cite{cirillo2002}, while the connected cluster  poses extra challenges and might change drastically the metastability scenarios.

As we have mentioned in Section~\ref{s:sincrono}, in this review we have not considered the case of PCA with non--null \emph{self--interaction}, that is 
to say the case in which $J_{00}\neq 0$ in equation \eqref{mod020} 
(i.e., the set $I$ contains the origin), since its dynamics cannot be written in terms of heat bath rates.  However, for $J_{00}\neq 0$
interesting and different metastability scenarios arise: the case $J_{00}=1$ has been studied in \cite{cirillo2008metastability} showing a similar metastable behavior to the Ising model with the asynchronous dynamics. 
For $0< J_{00}<1$ the paper \cite{cirillo2008competitive}
showed that, quite surprisingly, results similar to those found in \cite{cirillo1996metastability} for the Blume--Capel model are obtained.

As we have stressed in our manuscript, the synchronous dynamics converges to a measure that is different from the Ising Gibbs measure. Therefore, an interesting question is whether these two measure are close with respect to 
a suitable metric. A result in this direction is obtained in \cite{daipra2012}.  The role of the \emph{self--interaction} has been indeed further investigated, in order to derive an efficient method for approximating the sampling for the Ising stationary measure.  
In \cite[equation~(8)]{daipra2012} a PCA dynamics is introduced via
the \emph{lifted Hamiltonian}
\begin{equation}
\label{lifted}
H(\sigma,\eta)=
-\sum_{\newatop{i,j\in\Lambda}{i\neq j}} 
\bar{J}_{ij}\sigma(i)\eta(j)+\sum_{i\in\Lambda}q(1-\sigma(i)\eta(i))
\end{equation}
with $\bar{J}_{ij}$ a positive symmetric matrix and $q>0$ a positive parameter. 
The PCA dynamics is then the Markov chain defined by the transition
probabilities
\begin{equation}
\label{lifted02}
p_{T,q}(\sigma,\eta)=\frac{e^{-H(\sigma,\eta)/T}}{\sum_{\zeta\in\Omega} e^{-H(\sigma,\zeta)/T}}
\;.
\end{equation}
The parameter $q$ can be interpreted as an inertial term: for large values of $q$ the dynamics is very slow, flipping few spins at each time.
The transition probabilities (\ref{lifted02}) can be written as a product of single site updating probabilities, indeed,
\begin{equation}
\label{lifted03}
p_{T,q}(\sigma,\eta)=\prod_{i\in\Lambda}
\bar{f}_{T,\Theta_i\sigma}(\eta(i))
\end{equation}
with
\begin{equation}
\label{lifted04}
\bar{f}_{T,\Theta_i\sigma}(\eta(i))
=\frac{ 
\exp\Big\{\frac{1}{T}\eta(i)
\left[\sum_{j\in\Lambda\setminus\{i\}} 
\bar{J}_{ij}\sigma(j)+q\sigma(i)\right]\Big\}}
{2\cosh\Big\{\frac{1}{T}
\left[\sum_{j\in\Lambda\setminus\{i\}} 
\bar{J}_{ij}\sigma(j)+q\sigma(i)\right]\Big\}}
\;.
\end{equation}
In \cite{daipra2012} 
they posed for this model the question of Gibbsianess 
\cite{bertinicirilloolivieri1999,vanenterfernandezsokal1993},
it has been proved that the total variation distance 
between the invariant measure of the PCA defined in (\ref{lifted02})  
and the Ising Gibbs measure goes to zero as the volume and the 
control parameter $q$ goes to infinity. 
Notice that these approximation results do not 
apply directly to the model considered in the present review, 
which, indeed, is 
recovered for $q=h=0$ and $\bar{J}_{ij}=J_{ij}$. 
Afterwards, \cite{lancia2013}, \cite{daipra2015}  
extended the results of \cite{daipra2012} also to the case of \emph{weakly irreversible} PCA dynamics,  and \cite{procacci2016}  beyond the high temperature Dobrushin regime.

Finally, we mention 
the so called \emph{shaken} dynamics, which have been 
introduced in \cite{shaken2019} via a suitable lifted Hamiltonian 
having only down--left interactions and depending on a parameter $q$ 
that tunes the geometry of the system, that allows to interpolate 
between different lattices. For $q$  large the geometry is indeed the
square lattice, for finite $q$ the system lives on the hexagonal lattice, 
while for small
$q$ the system becomes the product of independent one dimensional lsing 
systems. Also for this dynamics the stationary measure tends to the 
Ising Gibbs measure in the thermodynamic limit. 

\bibliographystyle{unsrt}
\bibliography{cjs-review}

\end{document}